\documentclass[aps,prb,twocolumn,superscriptaddress,floatfix,preprintnumbers,showpacs]{revtex4-1}
\usepackage[caption=false]{subfig}
\usepackage{graphicx,mathtools,amssymb,color,mwe,soul}%,graphics
\usepackage{inputenc}
\DeclareMathOperator{\ef}{\mathcal{E}}
\DeclareMathOperator{\f}{\it{F}}
\newcommand{\unit}[1]{\,{\rm #1}}
\DeclareMathOperator{\lx}{\ell_\textit{x}}

\begin{document}

%\title{Comparison between $g$-tensor magnetic resonance and ``iso-Zeeman'' electric dipole spin resonance in hole quantum dots}

\title{Longitudinal and transverse electric field manipulation of hole spin-orbit qubits in one-dimensional channels}

\author{Vincent Philippe Michal}
\email{vincent.michal@cea.fr}
\affiliation{Univ. Grenoble Alpes, CEA, IRIG-MEM-L\_Sim, F-38000, Grenoble, France}
\author{Benjamin Venitucci}
\affiliation{Univ. Grenoble Alpes, CEA, IRIG-MEM-L\_Sim, F-38000, Grenoble, France}
\author{Yann-Michel Niquet}
\email{yniquet@cea.fr}
\affiliation{Univ. Grenoble Alpes, CEA, IRIG-MEM-L\_Sim, F-38000, Grenoble, France}

\begin{abstract}
Holes confined in semiconductor nanostructures realize qubits where the quantum mechanical spin is strongly mixed with the quantum orbital angular momentum. The remarkable spin-orbit coupling allows for fast all electrical manipulation of such qubits. We study an idealization of a CMOS device where the hole is strongly confined in one direction (thin film geometry), while it is allowed to move more extensively along a one-dimensional channel. Static electric bias and $ac$ electrical driving are applied by metallic gates arranged along the channel. In quantum devices based on materials with a bulk inversion symmetry, such as silicon or germanium, there exists different possible spin-orbit coupling based mechanisms for qubit manipulation. One of them, the $g$-tensor magnetic resonance ($g$-TMR), relies on the dependence of the effective $g$-factors on the electrical confinement. In this configuration the hole is driven by an $ac$ field parallel to the static electric field and perpendicular to the channel (transverse driving). Another mechanism, which we refer to here as iso-Zeeman electric dipole spin resonance (IZ-EDSR), is due to the Rashba spin-orbit coupling that leads to an effective time-dependent magnetic field experienced by the pseudo-spin oscillating along the quantum channel (longitudinal driving).
We compare these two modes of operation and we describe the conditions where the magnitudes of the Rabi frequencies are the largest. Different regimes can be attained by electrical tuning where the coupling to the $ac$ electric field is made either weak or strong. Spin-orbit coupling can also be tuned by strains, with, in particular, a transition from a mostly heavy- to a mostly light-hole ground state for in-plane tensile strains. Although large strains always reduce the Rabi frequency, they may increase the qubit lifetimes even faster, which calls for a careful optimization of strains and electric fields in the devices. We also discuss the choice of channel material and orientation. The study is relevant to the interpretation of the current experiments on the manipulation of hole qubits and as a guide to the development of quantum devices based on silicon and germanium.
\end{abstract}

\maketitle

\section{Introduction}

Spins in semiconductor quantum dots are envisioned as essential building blocks of future quantum processors and other quantum technologies\cite{Vandersypen2017, Loss1998, DiVincenzo2000, Hanson2007, Zwanenburg2013}. Their distinctive features include the possibility to be assembled in dense arrays of qubits, good coherence properties, and the ability to operate at relatively high temperatures \cite{Vandersypen2017, Yang2020, Petit2020}. Spin qubits in heterostructures made of silicon and germanium are particularly relevant because the materials can be isotopically purified. In such environments with negligible amount of nuclear spins, the coherence time of the qubits is greatly enhanced, and is ultimately limited by the quasi-stationary charge noise\cite{Yoneda2018}. 
Single and two-qubit operations of electronic spins in silicon have actually been demonstrated\cite{Pla2012, Veldhorst2014, Veldhorst2015, Kawakami2016, Yoneda2018, Watson2018, Zajac2018, Huang2019, Xue2019} with fidelities approaching the values compatible with fault-tolerant quantum computation.

Realizing qubits with holes instead of electrons can be attractive since in semiconductors such as silicon and germanium the spin-orbit interaction is much stronger in the valence than in the conduction band. This makes possible the all electrical manipulation of hole pseudo-spins\cite{comment_spin} without the need for micromagnets, and also the coupling of the effective spin with other degrees of freedom such as microwave photon modes in resonators\cite{Kloeffel2013}. 
Electrical manipulation of hole spin qubits has been shown experimentally in silicon metal-oxide-semiconductor (MOS) structures\cite{Maurand2016,Crippa2018} and in germanium\cite{Watzinger2018,Hendrickx2019}. Also in germanium arrays of hole quantum dots have been designed\cite{Lawrie2019,Scappucci2020,vanRiggelen2020} and multiple qubit logic has been demonstrated\cite{Hendrickx2020,Hendrickx2020_four_qubit}.

These advances motivate theoretical descriptions of the hole spin manipulation in cubic diamond materials such as silicon and germanium, which have an inversion symmetry center in bulk. The Rashba spin-orbit interaction has already been analyzed in nanowires\cite{Kloeffel2011,Kloeffel2018} and in planar (quasi-2D) geometries\cite{Bulaev2005, Bulaev2007, Marcellina2017, Terrazos2020}. The hole $g$-tensor modulation resonance ($g$-TMR) effect has also been described in connection with one-dimensional MOS channels on silicon-on-insulator (SOI)\cite{Venitucci2018, Venitucci2019}. In these structures, the time-dependent ($ac$) electric field that drags the hole is parallel to the static electric field that breaks the inversion symmetry of the dot. It modulates the $g$-factors of the dot and drives spin rotations. Ref. \citenum{Venitucci2019} does, in particular, include analytical and numerical calculations with the Luttinger-Kohn (LK) model in an idealized setup (almost identical to the one studied here but with a different confinement along the channel). The $g$-TMR Rabi frequency was derived in a minimal basis set and compared with the results of an exact diagonalization of the LK Hamiltonian in an extended basis. In Refs. \citenum{Venitucci2018} and \citenum{Venitucci2019}, the $g$-matrix formalism\cite{Kato2003} has proven useful in the numerical calculations of the Rabi frequency under electrical driving. 

Following these works, we investigate here an alternative way of manipulating the hole qubit with an $ac$ electric field perpendicular to the static electric field (and parallel to the channel). This $ac$ field drives the dot as a whole so that the Rashba spin-orbit interaction gives rise to an effective time-dependent magnetic field. This effect has been analyzed theoretically in Refs. \citenum{Rashba2003,Golovach2006} and we refer to it as iso-Zeeman electric-dipole spin resonance (IZ-EDSR \cite{Crippa2018}) because the Zeeman splitting of the qubit remains unchanged during the motion. We compare IZ-EDSR and $g$-TMR and we identify the regimes of operation where the Rabi frequencies are the largest. The spin-electric coupling can indeed be tuned by the static electric field and we show that the two effects are maximized in different conditions. Furthermore we study the influence of biaxial strain that can strongly change the interplay between the two mechanisms with a hole that transitions from a mostly heavy to a mostly light type at large enough tensile strain. With the perspective of optimizing the design we then discuss the material dependence and the influence of the device orientation.

The structure of the paper is as follows. In Section II we calculate the effective pseudo-spin Hamiltonian and the effective $g$-factors based on perturbation of the four-band LK Hamiltonian. In Section III, we recompute the $g$-TMR Rabi frequency with the g-matrix formalism, as an alternative derivation to Ref. 31. The latter, which is based on a power series expansion in a minimal basis set, includes some higher order contributions than the present work, but misses corrections on the effective $g$-factors due to the vector potential that are addressed here. We then discuss the conditions that optimize the $g$-TMR. In Section IV we analyze the IZ-EDSR starting from an effective Rashba spin-orbit coupling model and we also derive the conditions that maximize the Rabi frequency. In Section V we study the effect of strain and show how the situation changes when the qubit has a dominant light-hole character. In Section VI we discuss the results and compare the efficiency of $g$-TMR and IZ-EDSR. We also compare the analytical and semi-analytical results with numerical calculations based on the four-band LK model. Then we discuss the material dependence and the impact of the crystallographic orientation of the structure. We conclude in Section VII. In the Appendices we give details about the effective Hamiltonians (Appendix \ref{ap:EH}), the corrections to the $g$-factors that arise from the electromagnetic vector potential (Appendix \ref{ap:g_corr}), the derivation of the Rashba spin-orbit coupling model and the calculation of IZ-EDSR in one dimension (Appendices \ref{ap:Hso_para}, \ref{ap:EDSR} and \ref{ap:Hso_perp}), and we discuss additional figures in Appendix \ref{ap:add_figures}.

\section{Effective Zeeman Hamiltonian and $g$-tensor}\label{sec:HZ}

\begin{figure}[h!]
\centering
  \includegraphics[width=0.4\textwidth]{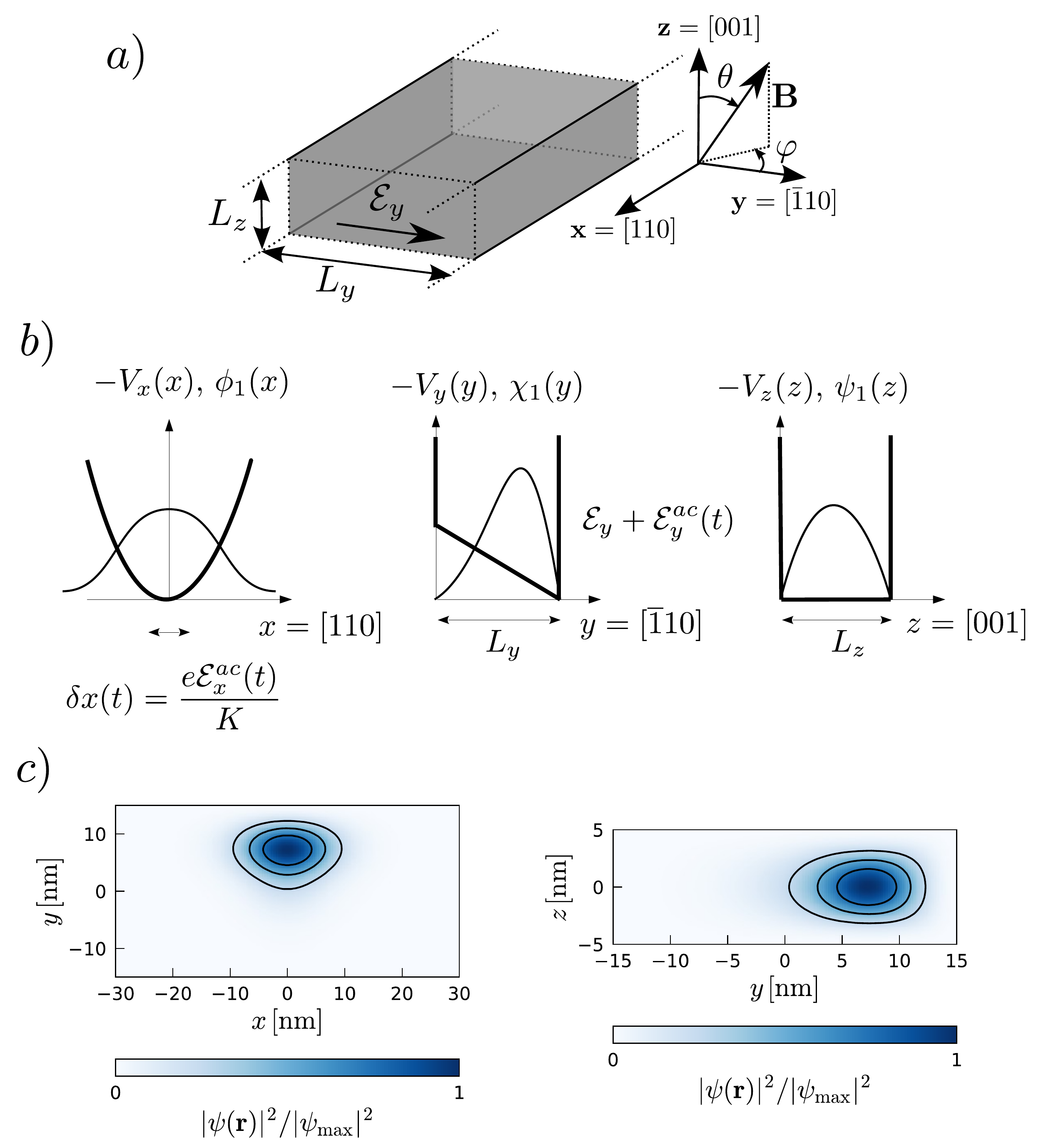}
\caption{The setup studied here: a) Three-dimensional perspective with system of coordinates and alignment with respect to the crystallographic axes. The structure has the smallest length $L_z$ along the direction of strong confinement $z=[001]$. b) Energy potential profiles along the $x$, $y$ and $z$ directions (see main text). A static electric field $\ef_y$ is applied along $y$. $ac$ electric fields are applied along $y$ and $x$ and lead to the $g$-TMR and IZ-EDSR effects respectively. c) Color maps of the probability densities in the $(xy)$ and the $(yz)$ planes that follow from the fundamental envelope functions numerically computed with the four-band LK model.\cite{Comment_numerics} In this calculation, $L_z=10\unit{nm}$, $L_y=30\unit{nm}$, the electric field is $\ef_y=0.5\unit{mV/nm}$, and $x_0\equiv(\pi\hbar)^{1/2}/(m_0K)^{1/4}=10\unit{nm}$ (note that $x_0$ is introduced here for convenience as a mass-independent characteristic length; the actual extent of the ground-state wave function of heavy- and light-holes along $x$ being given by Eq. (\ref{eq:ell_x})).}
\label{fig:setup}
\end{figure}

Motivated by spin qubit realizations in CMOS devices\cite{Maurand2016,Crippa2018} we consider a hole strongly confined along $z$ and weakly confined in the $(xy)$ plane. An idealization of the setup is shown in Fig. \ref{fig:setup}. We assume a rectangular channel along $x=[110]$ with hard wall boundary conditions and dimensions $L_y$ along $y=[1\overline{1}0]$ and $L_z\ll L_y$ along $z=[001]$ (infinite square well potentials along $y$ and $z$). A hole is confined along the channel in a parabolic potential $V_x(x)=-\frac{1}{2}K x^2$ (this setup slightly differs from Ref. \citenum{Venitucci2019} in order to allow for efficient spin manipulation with an $ac$ electric field along $x$). A static electric field $\ef_y$ (or equivalently a potential $V_y(y)=e\ef_y y$, $e>0$ being the elementary charge) is applied along $y$ that breaks the inversion symmetry of the channel and confines the hole towards the left or right facets. Additionally, a time-dependent ($ac$) electric field modulation is applied along $y$ or $x$. Note that we assume valence bands with negative dispersions, hence the signs of $V_x$ and $V_y$. 

In this quasi two-dimensional configuration and in the absence of strain the ground state is expected to have a dominant heavy-hole character with small mixing with light-hole envelopes (see for instance Ref. \citenum{Katsaros2011} for a summary of the properties of heavy-hole and light-hole states in a quasi-2D setup). We first analyze heavy-hole-like ground-states and we discuss the effects of strains and light-hole-like ground-states in Sec. \ref{sec:strain}. Because the mixing between the heavy-hole and the light-hole states is relatively small in the thin-film regime we derive effective quasi two-dimensional (quasi-2D) Hamiltonians by perturbation of the four-band LK model (defined in Ref. \citenum{Venitucci2019}). We describe the effective Hamiltonian method in Appendix \ref{ap:EH}. This approach is different from Ref. \citenum{Venitucci2019}, which solved the equations in a minimal basis set (and at higher order in perturbation) but missed some corrections on the hole masses and $g$-factors discussed in this work.

As shown in Ref. \citenum{Ares2013}, at leading order in the perturbation theory the heavy-hole/light-hole coupling leads to renormalization of the in-plane heavy-hole effective mass to
\begin{equation}\label{m_para_h}
m_\parallel^{h}=\frac{m_0}{\gamma_1+\gamma_2-\gamma_{h,1}},
\end{equation}
where the correction $\gamma_{h,1}$ reads\cite{Ares2013}:
\begin{equation}\label{gammah}
 \gamma_{h,1}=\frac{6\gamma_3^2\hbar^2}{m_0}\sum_{n}\frac{|\langle\psi_1^h|k_z|\psi_n^l\rangle|^2}{E_1^h-E_n^l}.
\end{equation}
Here $m_0$ is the bare electron mass, $\gamma_1$, $\gamma_2$, $\gamma_3$ are the Luttinger parameters characterizing the dispersion of the valence bands\cite{Comment_gammah}, $\psi_n^{h/l}$ are the envelopes of the heavy/light-hole states in the thin film, and $k_z=-i\partial/\partial_z$.
For a heavy-hole confined in an unstrained silicon quantum well the correction evaluates to $\gamma_{h,1}\approx 1.16$ while in germanium $\gamma_{h,1}\approx 3.56$.
As a consequence the in-plane envelope wavefunction of the heavy hole satisfies the 2D Schr{\"o}dinger equation with the effective mass $m_\parallel^h$. We show in Fig. \ref{fig:setup} b) the sketches of the envelope functions and in Fig. \ref{fig:setup} c) the color maps of the envelopes numerically computed with the four-band $k\cdot p$ (LK) model.  

The heavy-hole/light-hole coupling furthermore affects the $g$-tensor components. 
Without mixing between the heavy-hole and light-hole states the $g$-tensor of the heavy hole is diagonal\cite{Venitucci2018,Venitucci2019,Comment_g0} in the 
 $\{|J=3/2,J_z=3/2\rangle$, $|J=3/2,J_z=-3/2\rangle\}$ basis in use\cite{Comment_Heff}: 
\begin{equation}\label{g0}
g_0^h=\textrm{diag}(0, 0, -6\kappa).
\end{equation}
It was shown in Refs. \citenum{Ares2013} and \citenum{Watzinger2016} that at leading order in the perturbation theory the heavy-hole/light-hole coupling also leads to a renormalization of the $g$-factor in the direction of strong confinement:
\begin{equation}\label{gzh}
 g_z^{h}=-6\kappa+2\gamma_{h,1}.
\end{equation}

Moreover in the perturbation theory the effective Zeeman Hamiltonian acquires transverse components and writes (see Appendix \ref{ap:EH})
\begin{equation}\label{Heff}
H_{Z}^{h}=
\begin{pmatrix}
\frac{1}{2}g_z^{h}\mu_B B_z & \frac{2\sqrt{3}\langle R\rangle}{\Delta}\kappa\mu_B(B_x-iB_y) \\
\frac{2\sqrt{3}\langle R\rangle^*}{\Delta}\kappa\mu_B(B_x+iB_y) & -\frac{1}{2}g_z^{h}\mu_B B_z  
\end{pmatrix},
\end{equation}
with\cite{Venitucci2018, Venitucci2019, Comment_A}
\begin{equation}\label{R_Lutt}
R=\frac{\hbar^2}{2m_0}\sqrt{3}[-\gamma_3(k_x^2-k_y^2)+2i\gamma_2k_xk_y],
\end{equation}
that we average over the heavy-hole envelope function in the $(xy)$ plane, $k_a=-i\partial/\partial_a$ ($a=x,y$), and $\Delta=E_1^h-E_1^l$ is the energy gap between the topmost heavy-hole and light-hole states. For a hole strongly confined within the infinite well of width $L_z$ of the thin film, this energy gap is, neglecting strains and the influence of the split-off band: 
\begin{equation}\label{Delta}
\Delta=\frac{2\pi^2\gamma_2\hbar^2}{m_0L_z^2}.
\end{equation}
Thus Eq. (\ref{Heff}) gives the dependence of the effective $g$-factors on in-plane confinement:
\begin{subequations}
\begin{align}
 &g_{x}^h=g_{y}^h=-\frac{6\gamma_3\kappa\hbar^2}{m_0\Delta}\langle k_x^2-k_y^2\rangle,\label{geff_para}\\
 &g_{xy}^h=-g_{yx}^h=\frac{12\gamma_2\kappa\hbar^2}{m_0\Delta}\langle k_xk_y\rangle.
\end{align}
\end{subequations}
With the confinement considered here the off-diagonal element $g_{xy}$ vanishes \cite{Comment_A}.

\begin{figure}
\centering
  \includegraphics[width=0.4\textwidth]{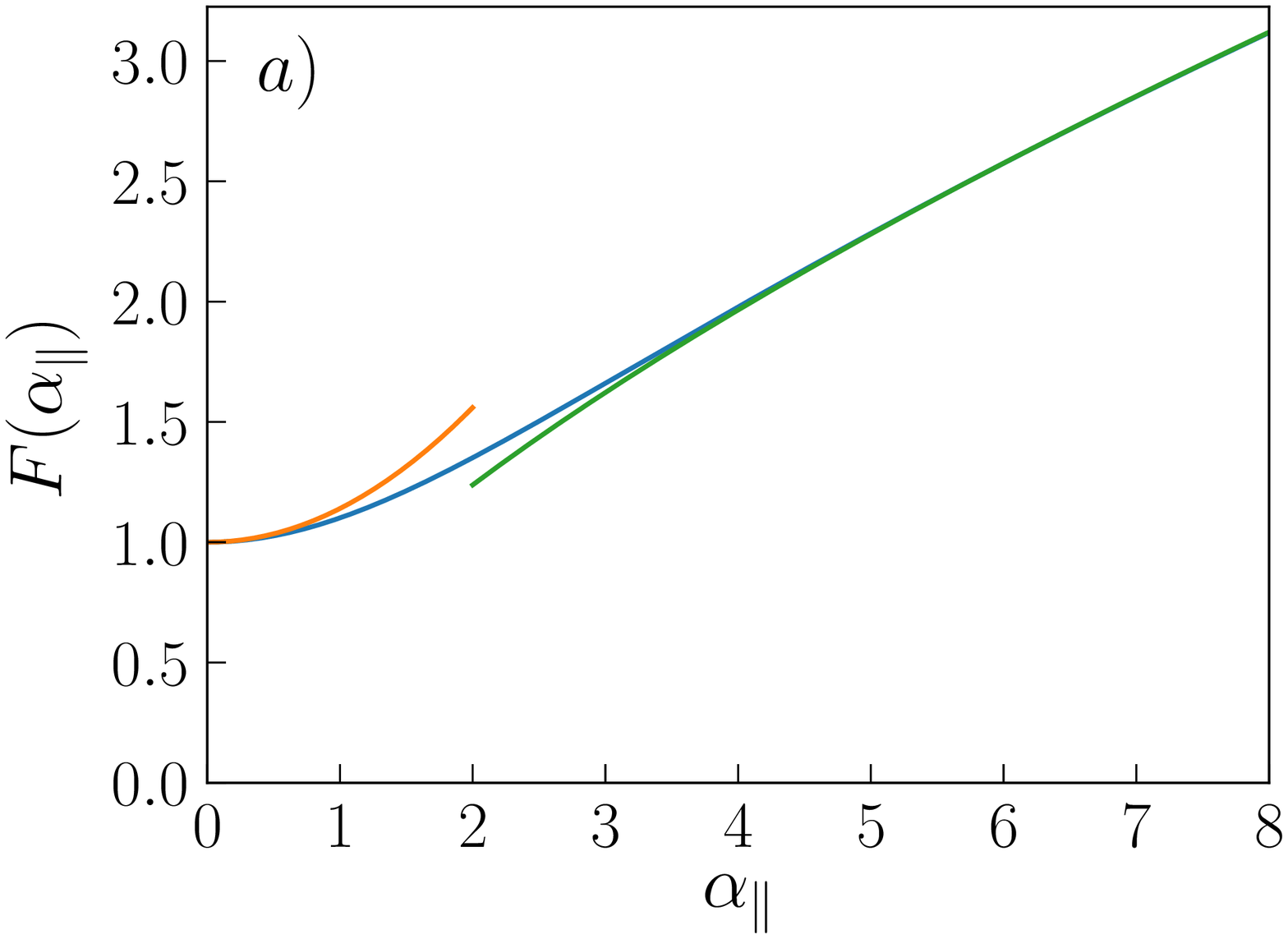}\\
  \includegraphics[width=0.4\textwidth]{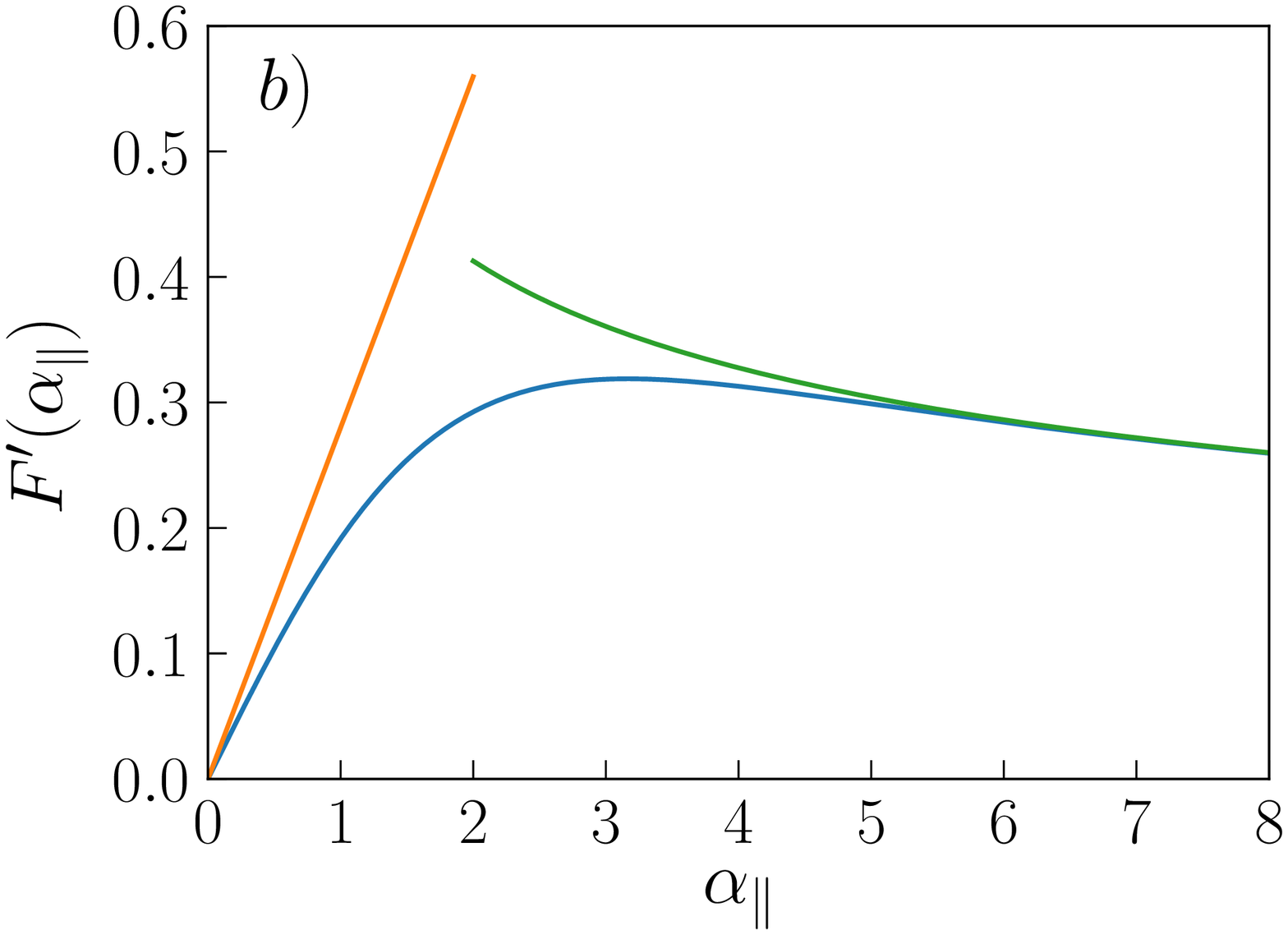}\\
\caption{a) The function $\f(\alpha_\parallel)=\langle k_y^2\rangle L_y^2/\pi^2$ and b) its derivative $\f'(\alpha_\parallel)$ where $\alpha_\parallel$ is the dimensionless parameter defined by Eq. (\ref{alpha_pl}). The blue curve corresponds to the numerical calculation, the orange and the green curves correspond to the weak electric field and to the strong electric field asymptotics [Eq. (\ref{asymp})] respectively.}
\label{fig_ffp}
\end{figure}

The hole motion is separable in the ($xy$) plane ; the eigensolutions along $x$ are those of the 1D harmonic oscillator, and we solve numerically the Schr\"odinger equation along $y$ with the method of Fourier series\cite{Venitucci2019}. We introduce the scaling function
\begin{equation}\label{sf}
 \f(\alpha_\parallel)=\frac{\langle k_y^2\rangle L_y^2}{\pi^2},
\end{equation}
where we take the average over the ground state envelope function. The parameter $\alpha_\parallel$ quantifies the relative importance of structural and electric confinements:
\begin{equation}\label{alpha_pl}
\alpha_\parallel=\frac{2m_\parallel	e\ef_yL_y^3}{\pi^3\hbar^2}=\Big(\frac{L_y}{\pi\ell_{\ef_y}}\Big)^3.
\end{equation}
In the above equation $m_\parallel$ can be the effective mass of a heavy hole or a light hole (the latter case will be discussed in Sec. \ref{sec:strain}). Also $\ell_{\ef_y}=(\hbar^2/(2m_\parallel e\ef_y))^{1/3}$ is the characteristic length of confinement by the electric field $\ef_y$. The asymptotics of the scaling function (\ref{sf}) are
\begin{equation}\label{asymp}
\f(\alpha_\parallel)\approx\left\{
\begin{array}{rl}
1+c_1\alpha_\parallel^2\text{, } \alpha_\parallel\ll 1,\\
\frac{|a_1|}{3}\alpha_\parallel^{2/3}\text{, } \alpha_\parallel\gg 1,
\end{array} \right.
\end{equation}
with $c_1\approx0.14$ and $a_1\approx-2.34$ is the first zero of the Airy function ${\rm Ai}$. Plots of $\f$, its derivative, and comparison with the asymptotics Eq. (\ref{asymp}) are shown in Fig. (\ref{fig_ffp}). We note $g_\parallel^h=g_{x}^h=g_{y}^h$, and introduce the extent of the wave function along $x$:
\begin{equation}\label{eq:ell_x}
\ell_x^2=2\langle x^2\rangle=\frac{\hbar}{(m_\parallel K)^{1/2}}=\frac{1}{2\langle k_x^2\rangle},
\end{equation}
to get:
\begin{equation}\label{eq:g_para}
g_\parallel^h=\frac{6\gamma_3\kappa\hbar^2}{m_0\Delta}\Big(\frac{\pi^2\f(\alpha_\parallel^h)}{L_y^2}-\frac{1}{2\lx^2}\Big).
\end{equation}
The effective in-plane $g$-factor hence depends on the in-plane electric field. This can lead to spin coherent oscillations under $ac$ electrical driving\cite{Kato2003,Venitucci2018,Venitucci2019} as we show in the next section.

\section{$g$-tensor magnetic resonance}\label{sec:gtmr}

The $g$-TMR mechanism has been recently analyzed numerically and analytically in Refs. \citenum{Venitucci2018} and \citenum{Venitucci2019}. The present set-up, with the $ac$ electric field applied along $y$, is a paradigm of this mechanism. It is practically realized when the same gate partly overlapping the channel is used to apply the static electric field $\ef_y$ and the $ac$ modulation $\ef_y^{ac}$. The Rabi oscillations then result from the electrical modulation of the principal $g$-factors $g_\parallel^h$ and $g_z^h$ in the anharmonic confinement potential $V_y(y)$ shaped by the structural confinement and transverse electric field $\ef_y$ \cite{Venitucci2018}. Here we give an alternative analytical derivation of the Rabi frequency based on the $g$-matrix formalism\cite{Kato2003,Venitucci2018,Venitucci2019}, and we discuss additional corrections that come from the vector potential terms derived in Appendix \ref{ap:g_corr}.
In this formalism the Rabi frequency is computed at linear order in the applied magnetic field and in the $ac$ gate voltage. Because the box that contains the hole behaves as a parallel plate capacitor, we can express the Rabi frequency in term of the electric fields instead of gate voltages: 
\begin{equation}\label{fRg}
f_{Rg} = \frac{\mu_B\ef_y^{ac}\|(g{\bf B})\times(\frac{\partial g}{\partial\ef_y}{\bf B})\|}{2h\| g{\bf B}\|},
\end{equation}
%Here $g'$ the derivative of the $g$-matrix with respect to $\ef_y$.
with $h$ the Planck constant and $\ef_y^{ac}$ the amplitude of the $ac$ electric field along $y$: $\ef_y^{ac}(t)=\ef_y^{ac}\sin(\omega t)$. The qubit is resonantly driven at the average Larmor angular frequency $\omega_L = \mu_B\|g{\bf B}\|/\hbar$.
With the $g$-matrix $g = \textrm{diag}(g_\parallel^h, g_\parallel^h, g_z^h)$ the Rabi frequency becomes %and its derivative with respect to $\ef_y$, $\frac{\partial g^h}{\partial\ef_y} = \textrm{diag}(\frac{\partial g_\parallel^h}{\partial\ef_y}, \frac{\partial g_\parallel^h}{\partial\ef_y},\frac{\partial g_z^h}{\partial\ef_y})$, 
\begin{equation}
f_{Rg}^h= \frac{\mu_B \ef_y^{ac}|\frac{\partial g_\parallel^h}{\partial\ef_y}g_z^hB_\parallel B_z|}{2h\sqrt{(g_\parallel^h B_\parallel)^2+(g_z^h B_z)^2}},
\end{equation}
with $B_\parallel=\sqrt{B_x^2+B_y^2}$. In the thin film regime the Rabi frequency has an approximate rotational symmetry with respect to the magnetic field orientation in the $(xy)$ plane\cite{Venitucci2019}. On the other hand it strongly depends on the angle $\theta$ between the magnetic field and the $z$ axis\cite{AresAPL,Venitucci2019} and reaches a maximum 
\begin{equation}\label{fRgmax}
f_{Rg\max}^h= \frac{\mu_B B\ef_y^{ac}|\frac{\partial g_\parallel^h}{\partial\ef_y}|}{2h(|g_\parallel^h/g_z^h|+1)}
\end{equation}
at the angles  
$
\theta_{\max}=\pi/2\pm\arctan\Big(\sqrt{|g_\parallel^h/g_z^h|}\Big).
$
Holes with dominant heavy character fulfill $g_\parallel^h\ll g_z^h$ and the optimal angles approximate as
\begin{equation}\label{thetamax}
 \theta_\textrm{max}\approx \pi/2\pm\sqrt{|g_\parallel^h/g_z^h|}.
\end{equation}

With Eqs. (\ref{sf}) and (\ref{alpha_pl}) the Rabi frequency of a heavy hole develops as
\begin{equation}\label{fRgmaxh}
 f_{Rg\max}^h\approx\frac{6\gamma_3|\kappa|\mu_B Be\ef_y^{ac}L_y\f'(\alpha_\parallel^h)}{\pi h(m_0/m_\parallel^h)\Delta},
\end{equation}  
together with the asymptotics of the derivative:
\begin{equation}
\f'(\alpha_\parallel^h) \approx
\begin{dcases}
\frac{4c_1m_\parallel^h e\ef_yL_y^3}{\pi^3\hbar^2},&L_y/\pi\ll\ell_{\ef_y},\\
\frac{2\pi|a_1|\ell_{\ef_y}}{9L_y},&L_y/\pi\gg\ell_{\ef_y}.
\end{dcases}
\end{equation}
As a function of $\ef_y$ it reaches a maximum 
\begin{equation}\label{fRgmaxs}
 f_{Rg\max*}^h\approx\frac{0.6\gamma_3|\kappa|\mu_B Be\ef_y^{ac}L_y}{h(m_0/m_\parallel^h)\Delta}
\end{equation}
at $e\ef_{y*}\approx1.25\pi^3\hbar^2/(m_\parallel^h L_y^3)$, which is consistent with Eqs. (42), (43) of Ref. \citenum{Venitucci2019}, given the different approximations made here and in Ref. \citenum{Venitucci2019}. For a heavy-hole spin qubit in silicon with $B=1\unit{T}$, $\ef_y^{ac}L_y\sim1\unit{mV}$, and $\Delta\sim5\unit{meV}$ ($L_z=10\unit{nm}$), this evaluates as $f_{Rg\max*}^h\sim 250\unit{MHz}$. 

\begin{figure*}
\centering
  \includegraphics[width=0.4\textwidth]{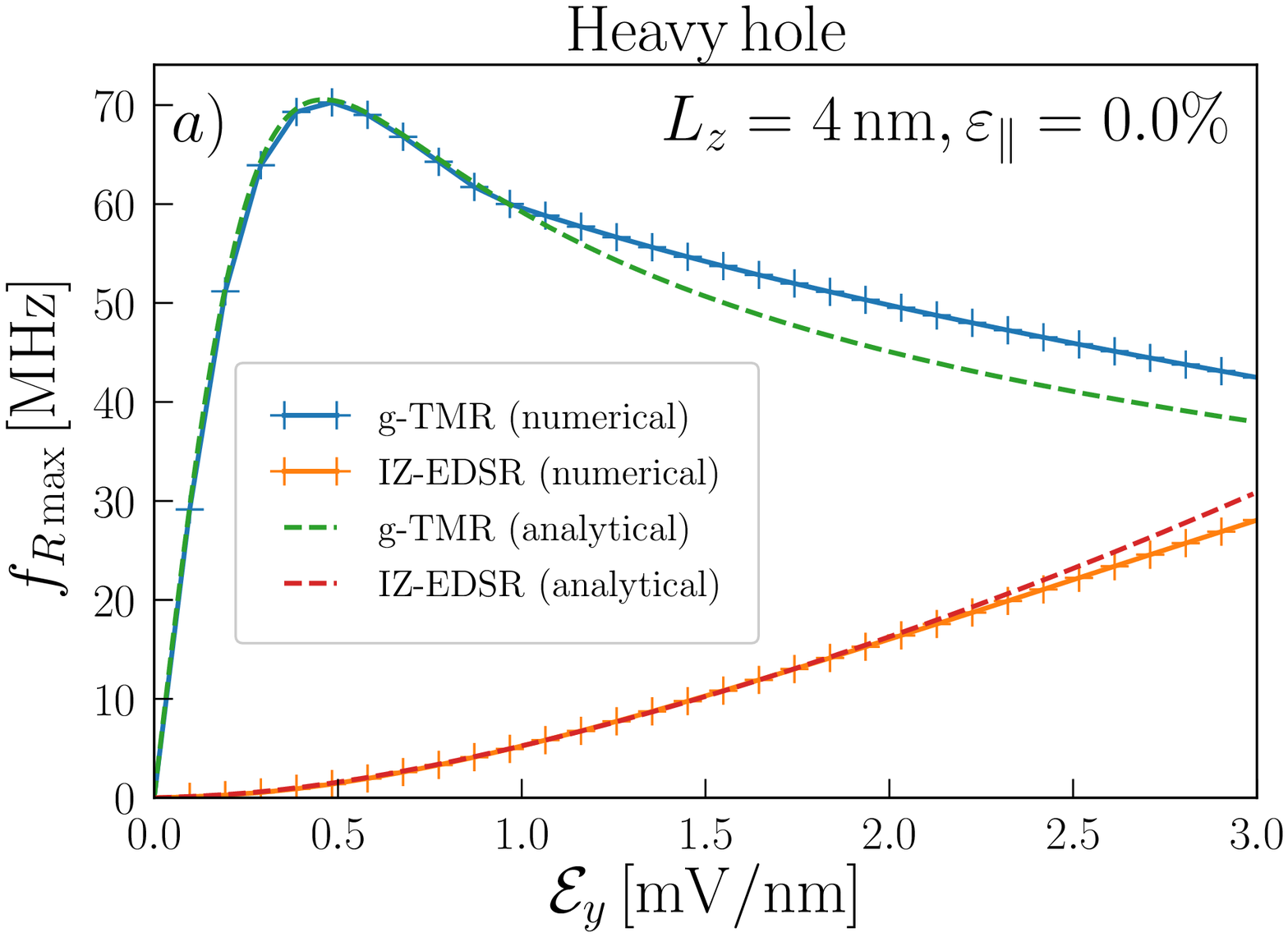}
  \includegraphics[width=0.4\textwidth]{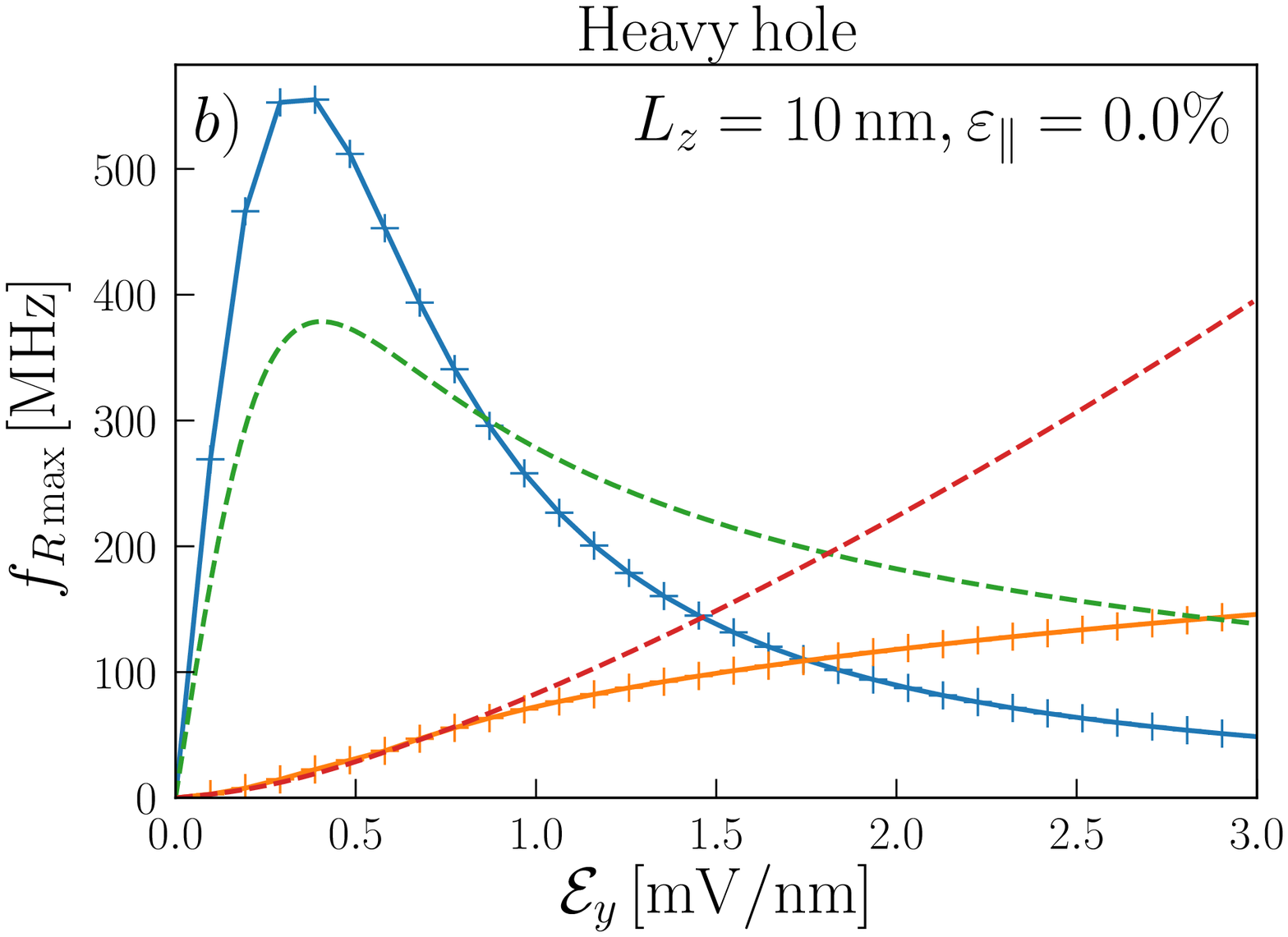}\\
  \includegraphics[width=0.4\textwidth]{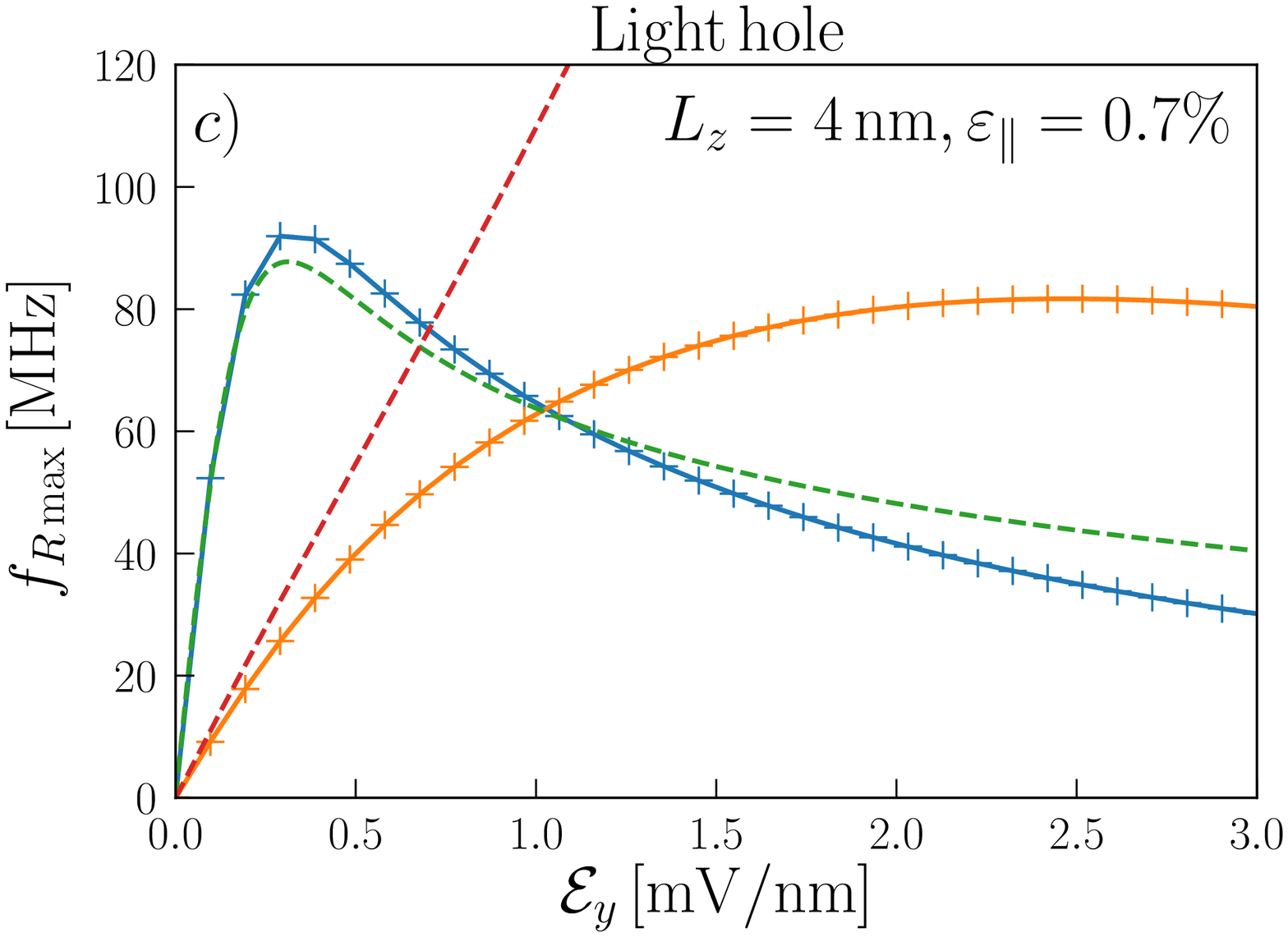}
  \includegraphics[width=0.4\textwidth]{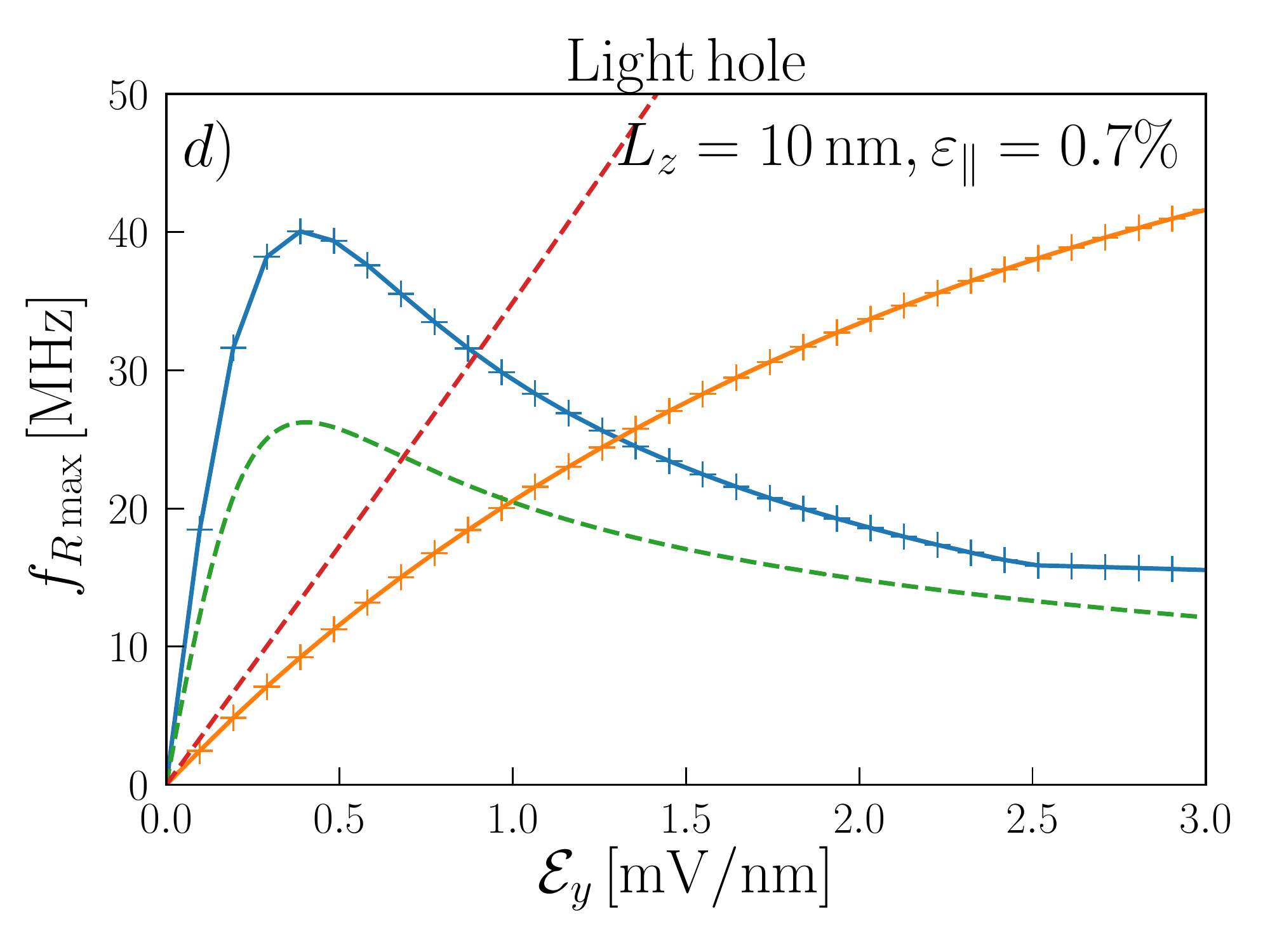}
\caption{Comparison of the maximal $g$-TMR and the IZ-EDSR Rabi frequencies as given by Eq. (\ref{fRg}), the $g$-factors Eqs. (\ref{g_h_corr}), (\ref{g_l_corr}), and Eq. (\ref{fRipara}) (dashed lines), with numerical calculations based on the four band $k\cdot p$ model \cite{Comment_numerics} (full lines). The material is silicon and the parameters are $\varepsilon_\parallel=0\%$ for Figures a) and b), $\varepsilon_\parallel=0.7\%$ for Figures c) and d),  $L_z=4\unit{nm}$ for Figures a) and c), $L_z=10\unit{nm}$ for Figures b) and d), $B=1\unit{T}$, $\ef_{x/y}^{ac}=(1/30)\unit{mV/nm}$, $L_y=30\unit{nm}$, and $x_0=10\unit{nm}$.}
\label{fig:comparison}
\end{figure*}

There are corrections beyond Eq. (\ref{eq:g_para}) that break the rotational symmetry of the $g$-tensor in the $(xy)$ plane ($g_x^h\neq g_y^h$). In Ref. \citenum{Venitucci2019} such corrections arose in the perturbation series at higher orders in the parameter $L_z/L_y\ll 1$.
We derive in Appendix \ref{ap:g_corr} other anisotropic corrections due to the electromagnetic vector potential (whose action was neglected in Ref. \citenum{Venitucci2019}). With the corrected $g$-factors we compute the maximal Rabi frequencies semi-analytically with Eq. (\ref{fRg}). In Fig. \ref{fig:comparison} we compare the semi-analytical results with the fully numerical calculations based on the exact solution of the four-band $k\cdot p$ model that follows from previously developed methods\cite{Venitucci2019,Comment_numerics}. 
We further comment Fig. \ref{fig:comparison} in Section \ref{sec:discussion}, where we will also make the comparison with the IZ-EDSR effect that we describe in the next section. 

\section{Iso-Zeeman EDSR}\label{sec:iso_para}

The $ac$ electric field may be aligned with the channel (along $x$) rather than perpendicular to it (along $y$). As the confinement is parabolic along $x$, such a modulation drags a real-space oscillation of the dot as a whole with amplitude
\begin{equation}\label{dx}
\delta x=\frac{e\ef_x^{ac}}{K}=\frac{e\ef_x^{ac}m_\parallel\lx^4}{\hbar^2},
\end{equation}
where $\ef_x^{ac}$ is the amplitude of the oscillating electric field $\ef_x^{ac}(t)=\ef_x^{ac}\sin(\omega t)$. The qubit is resonantly driven so that the angular frequency $\omega$ is set to the Larmor frequency of the effective two-level system. Then the spin-orbit interaction leads to an effective magnetic field that is position-dependent \cite{Aleiner2001,Levitov2003} and the oscillating hole experiences an effective time-dependent magnetic field that can lead to coherent oscillations of the (pseudo) spin \cite{Rashba2003,Golovach2006,Nowack2007}.

\begin{figure}[h]
\centering
  \includegraphics[width=0.4\textwidth]{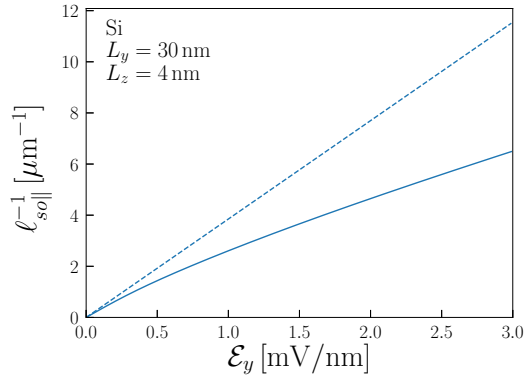}
\caption{Inverse spin-orbit length as a function of the static electric field along $y$. We compare the semi-analytical formula Eq. (\ref{l_so_inv}) of Appendix \ref{ap:Hso_para} (solid line) with the linear approximation Eq. (\ref{l_so_inv_lin}) (dashed line) for a silicon channel with dimensions $L_z=4\unit{nm}$ and $L_y=30\unit{nm}$.}
\label{fig:lso}
\end{figure}

In the typical gate configuration of Refs. \citenum{Maurand2016,Crippa2018,Venitucci2018}, the effective Rashba spin-orbit coupling is mostly ruled by the in-plane static electric field $\ef_y$ because the electrical polarizability is much weaker along $z$ due to the strong confinement (the effect of a static electric field $\ef_z$ will be briefly discussed in section \ref{sec:discussion}). With the method presented in Appendix \ref{ap:EH} we derive the Rashba Hamiltonian (see details of the calculation in Appendix \ref{ap:Hso_para}):
\begin{equation}\label{HR1D_para}
 H_{so\parallel}=\frac{\hbar^2}{m_\parallel\ell_{so\parallel}}k_x\sigma_z.
\end{equation}
The basis employed is as before and we have used the Pauli matrix notation in order to express the Hamiltonian in a compact form. 
In the thin film limit and for $L_y/\pi<\ell_{\ef_y}$ the inverse effective spin-orbit length given by Eq. (\ref{l_so_inv}) of Appendix \ref{ap:Hso_para} is well approximated by:
\begin{equation}\label{l_so_inv_lin}
\ell_{so\parallel}^{-1}\approx\frac{3\gamma_2\gamma_3e\ef_y}{(m_0/m_\parallel)^2\Delta}.
\end{equation}
In Fig. \ref{fig:lso} we plot the inverse spin-orbit length computed with Eq. (\ref{l_so_inv}) and we compare it with the approximation Eq. (\ref{l_so_inv_lin}). 

The Rashba spin-orbit interaction Eq. (\ref{HR1D_para}) implies the time-dependent effective Zeeman interaction \cite{Golovach2006} (see Appendix \ref{ap:EDSR})
\begin{equation}\label{HZt}
\delta H_Z(t)=\frac{\delta x}{\ell_{so\parallel}}\sin(\omega t)\mu_B(g_x B_x\sigma_y-g_y B_y\sigma_x),
\end{equation} 
where $\delta x$ is given by Eq. (\ref{dx}). 
The time-dependent effective magnetic field associated to Eq. (\ref{HZt}) is perpendicular to the external magnetic field\cite{Golovach2006} and immediately yields the Rabi frequency 
\begin{equation}\label{fRipara}
f_{Ri\parallel}=\frac{\delta x}{h\ell_{so\parallel}}\mu_B\sqrt{g_x^2 B_x^2+g_y^2 B_y^2}.
\end{equation}
The result does not depend on the character of the hole and holds for a mostly heavy as well as a mostly light hole. The light-hole case will be addressed in Sec. \ref{sec:strain}. For the heavy hole the Rabi frequency Eq. (\ref{fRipara}) is approximately symmetric with respect to the magnetic field orientation in the $(xy)$ plane and reaches a maximum when the magnetic field is in the equatorial plane ($B_z=0$):
\begin{equation}\label{fRiparamax}
f_{Ri\parallel\max}^h=\frac{\delta x}{h\ell_{so\parallel}}\max(|g_x^h|,|g_y^h|)\mu_B B,
\end{equation}
where $g_x^h$ and $g_y^h$ are given by Eq. (\ref{g_h_corr}).

When the electric field is so strong that $\ell_{\ef_y}<L_z/\pi$, but still $\Delta<\Delta_{so}$ ($\Delta_{so}$ being the spin-orbit energy gap between the split-off bands and the heavy- and light-hole bands), then the Rashba spin-orbit coupling can be addressed in the quasi-two-dimensional regime with strong confinement in the direction of the electric field.  
We numerically find (see Fig. \ref{fig:loglog}a of Appendix \ref{ap:add_figures}) that for a strong electric field the Rabi frequency decreases as $\ef_y^{-1/3}$. Qualitatively, the energy separation between the confined states is now dominated by the electric field ($\propto\ell_{\ef_y}^{-2}$) and the matrix elements responsible for the spin-orbit coupling are linear in the momentum in the direction of the strong confinement ($\propto\ell_{\ef_y}^{-1}$). Since the $g$-factors saturate to constants for strong electric fields (see Fig. \ref{fig:loglog}b of Appendix \ref{ap:add_figures}), the Rabi frequency must be asymptotically proportional to $\ell_{\ef_y}\propto\ef_y^{-1/3}$ according to Eq. (\ref{fRipara}).

The Rabi frequency is therefore maximum in the range where the energies of confinement in the $y$ and $z$ directions are comparable.
This intermediate regime is in fact similar to the nanowire (quasi-1D) configuration\cite{Kloeffel2018} where the Rashba spin-orbit coupling remains of the form of Eq. (\ref{HR1D_para}) with an inverse spin-orbit length that expresses as:	
\begin{equation}
\ell_{so\parallel}^{-1}=C\frac{e\ef_y}{\Delta},
\end{equation}
$\Delta$ being the energy splitting between the two relevant Kramers pairs of the four-band LK model, and $C$ a dimensionless factor that depends on the LK parameters. 
We have numerically computed the maximum EDSR Rabi frequencies for silicon and germanium. 
They are reached for a magnetic field oriented in the $y$-direction since the component of the $g$-tensor with the largest magnitude is $g_y$ in this regime (see Fig. \ref{fig:loglog}b of Appendix \ref{ap:add_figures}). For silicon with $L_z=10\unit{nm}$ and in the absence of strain, the Rabi frequency tends to saturate when $\ef_{y*}\gtrsim 10$ mV/mn and reaches a maximum $f_{Ri\max*}\sim270\unit{MHz}$ at $\ef_{y*}\sim40\unit{mV/nm}$. This large field is, however, practically beyond the operating range of CMOS qubits (and actually above the breakdown field of bulk silicon).

In the present and in the previous sections we have analyzed the IZ-EDSR and the $g$-TMR as two distinct mechanisms. However we remind that IZ-EDSR can be accompanied by a $g$-TMR-like contribution if the confinement potential along $x$ is not strictly parabolic\cite{Crippa2018}. 

\begin{figure*}
\centering
  \includegraphics[width=0.4\textwidth]{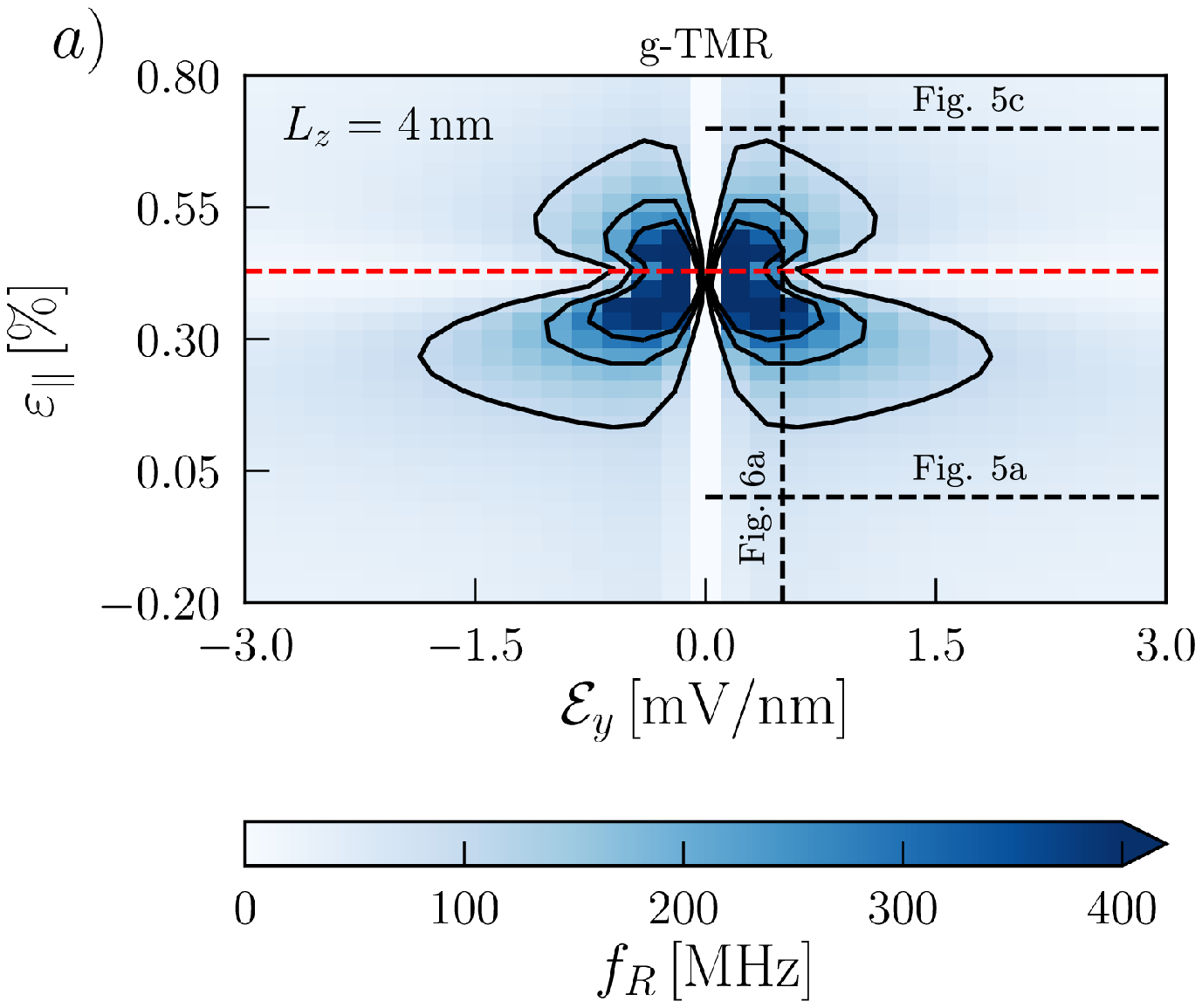}
  \includegraphics[width=0.4\textwidth]{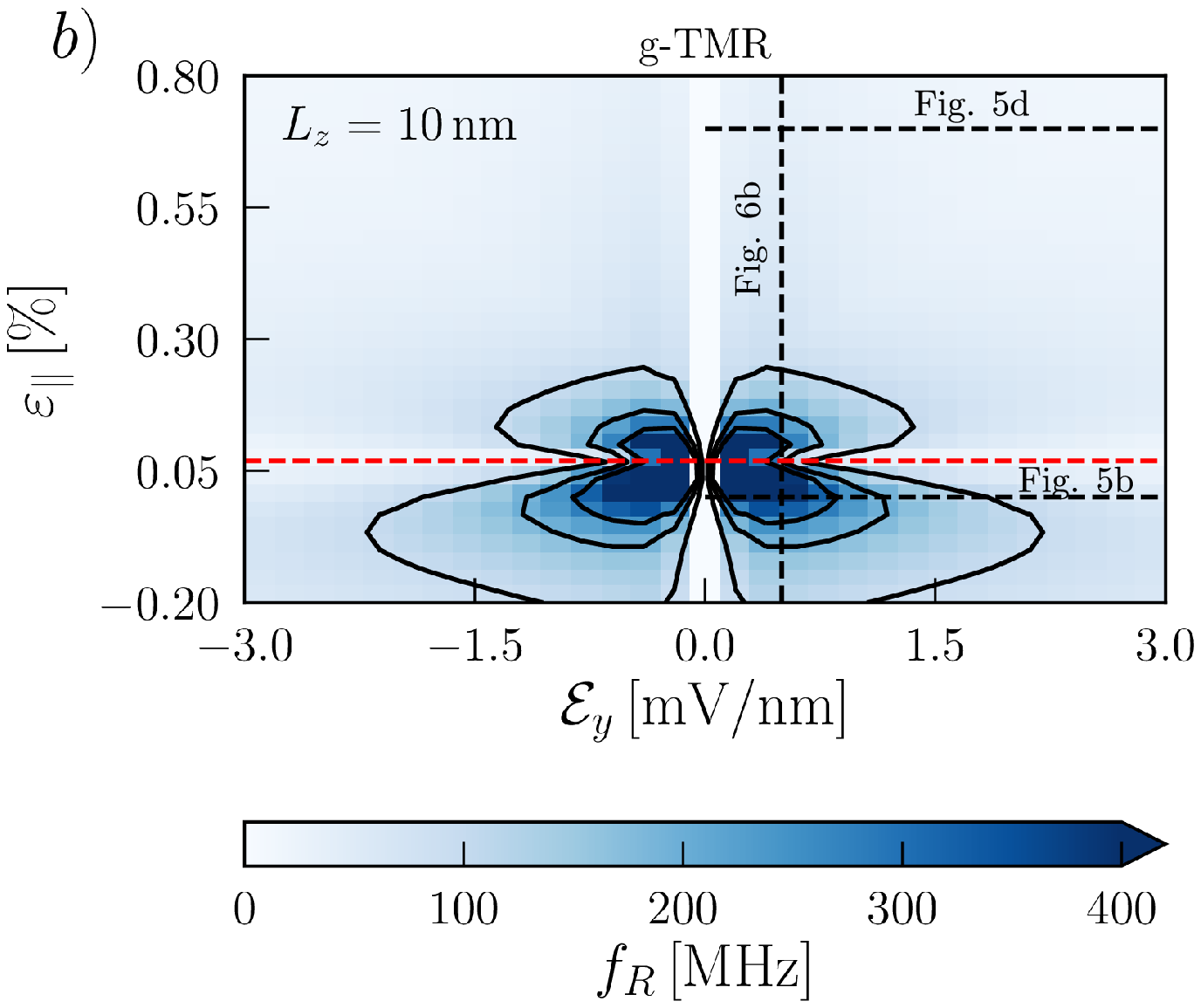}
  \includegraphics[width=0.4\textwidth]{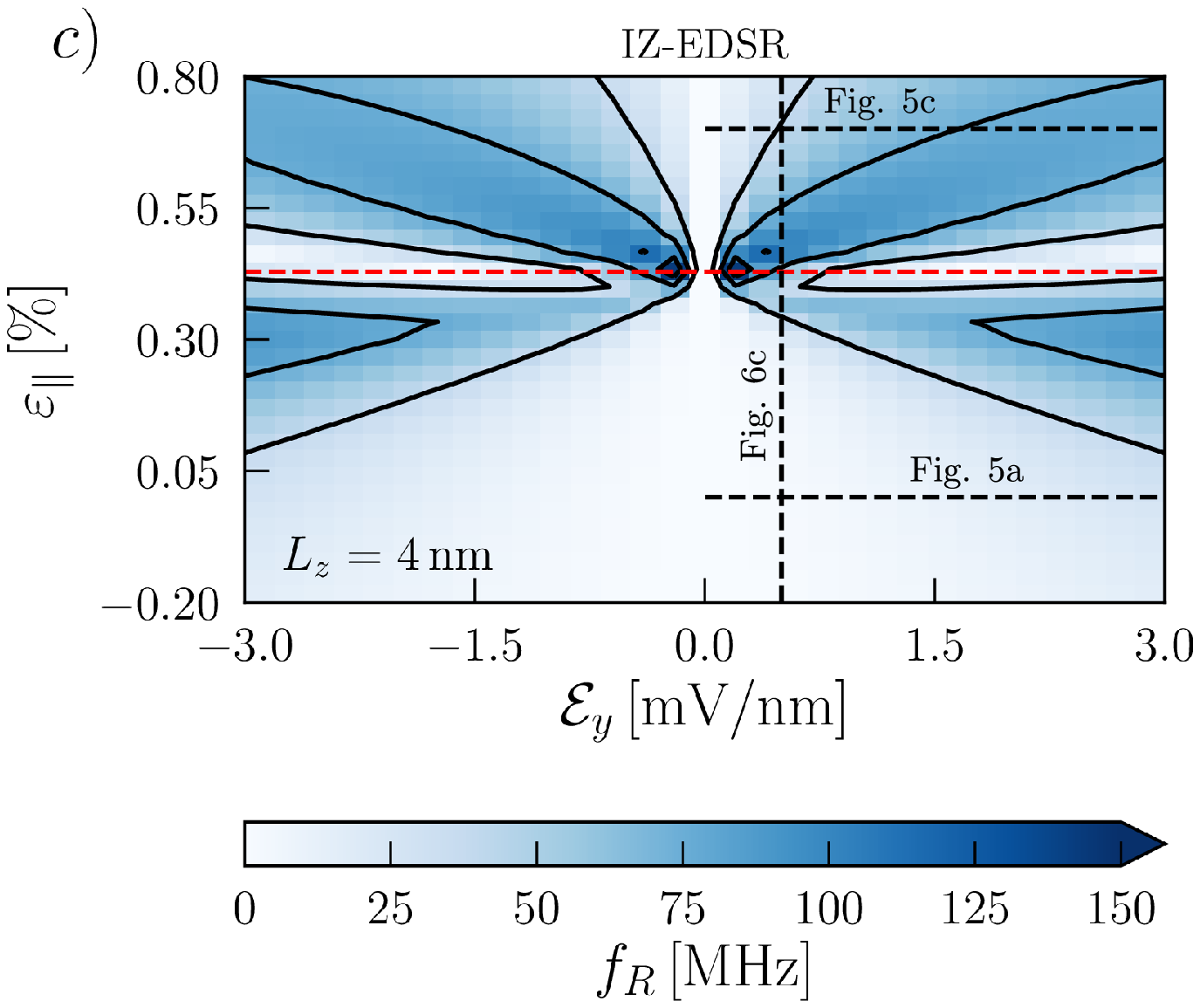}
  \includegraphics[width=0.4\textwidth]{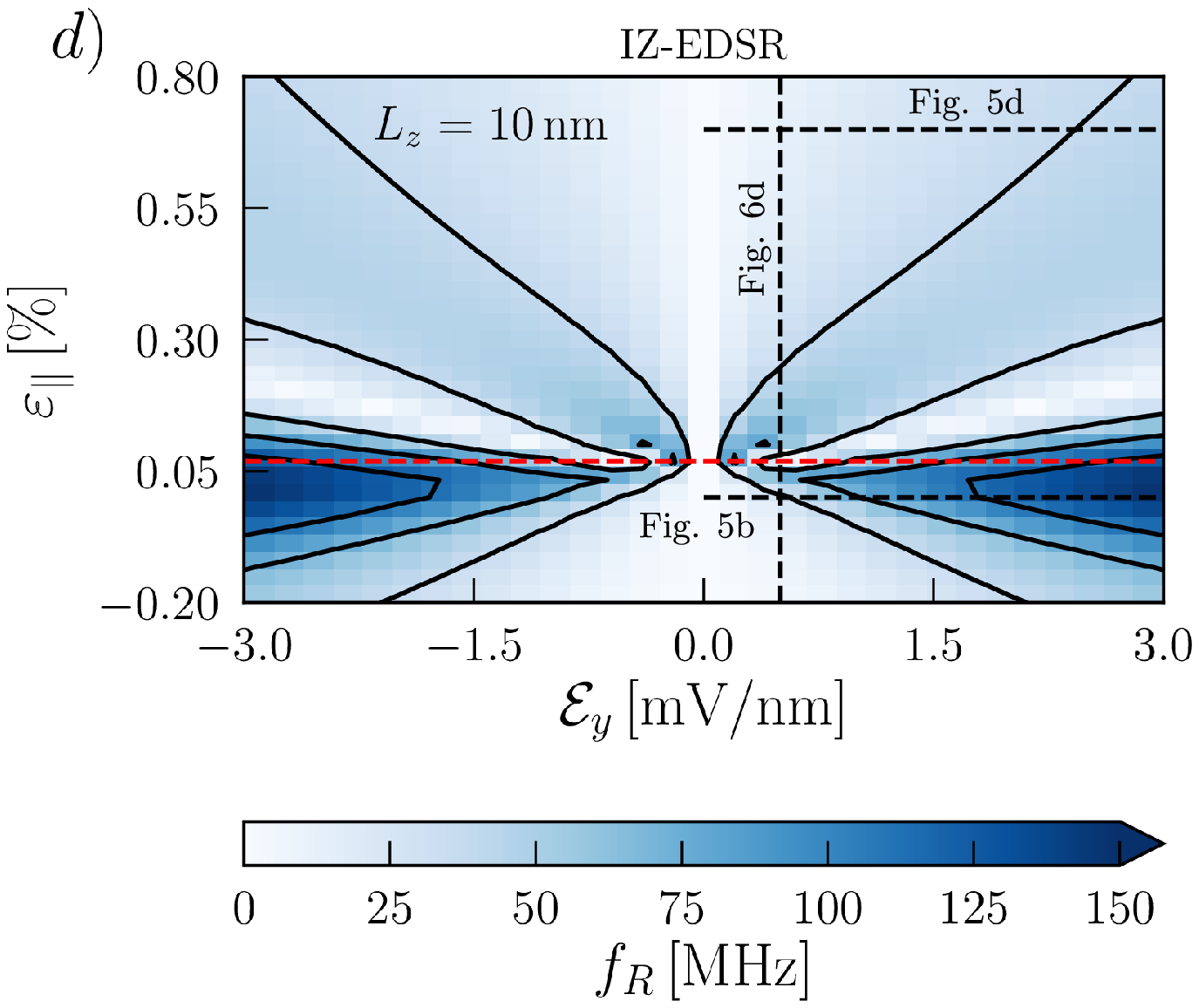}
\caption{Maps of the Rabi frequency for the $g$-TMR effect (maps a) and b)) and the IZ-EDSR effect (maps c) and d)), as a function of the lateral electric field $\ef_y$ and biaxial strain $\varepsilon_\parallel$. The maps are obtained from a numerical solution of the four-band $k\cdot p$ model\cite{Comment_numerics} in silicon. The lengths that characterize the lateral confinement are $x_0\equiv(\pi\hbar)^{1/2}/(m_0K)^{1/4}=10\unit{nm}$ and $L_y=30\unit{nm}$.
The height of the semiconductor channel is $L_z=4\unit{nm}$ for maps a) and c), and it is $L_z=10\unit{nm}$ for maps b) and d). The dashed black lines outline the constant strains and electric field cuts shown in Figs. \ref{fig:comparison} and \ref{fig:strain}. The red dashed lines mark the critical strain $\varepsilon_{\parallel}^{\ast}$ that separates the mostly heavy-hole from the mostly light-hole ground state.}
\label{fig:map}
\end{figure*}

\begin{figure*}
\centering
  \includegraphics[width=0.4\textwidth]{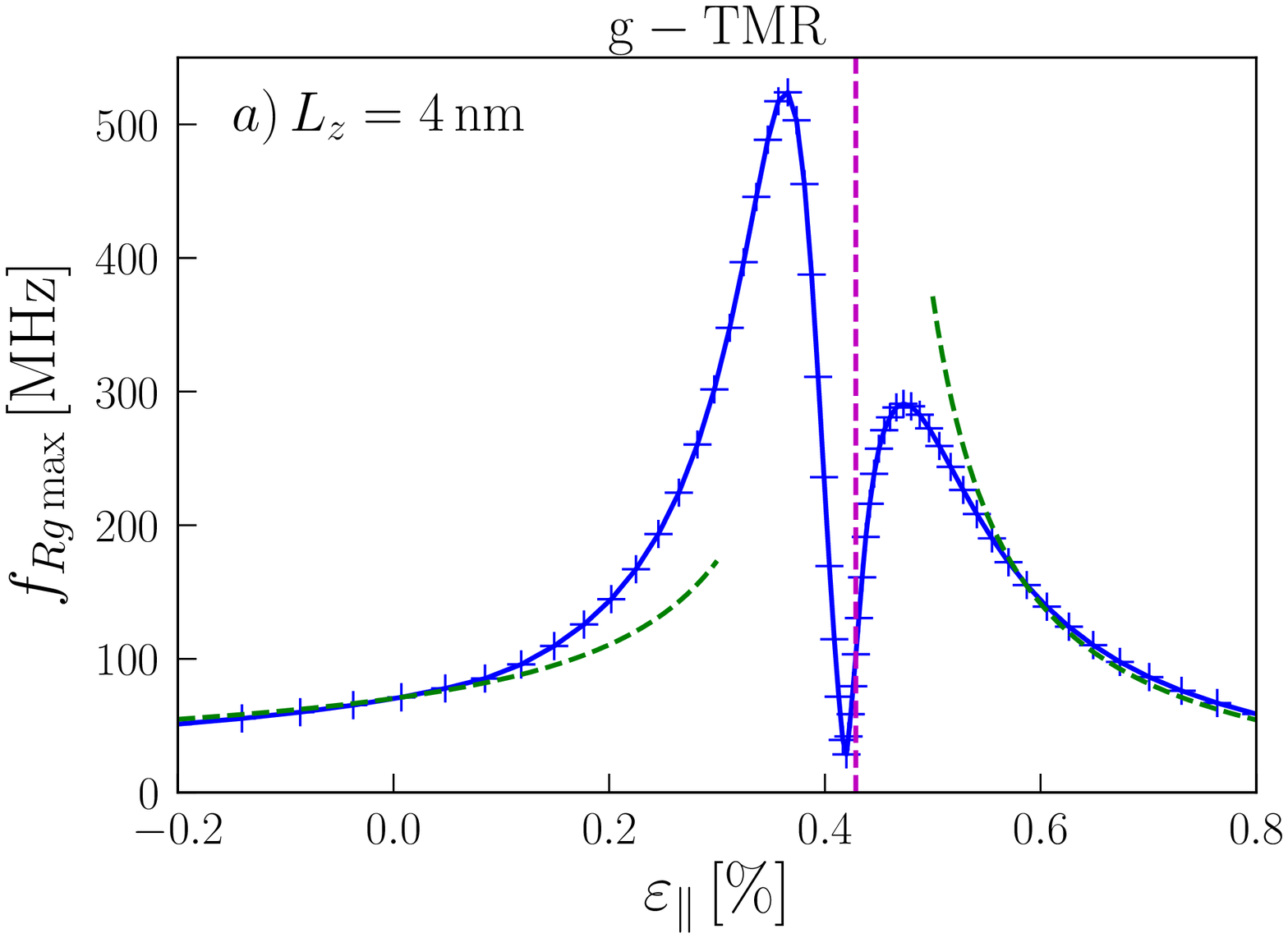}
  \includegraphics[width=0.4\textwidth]{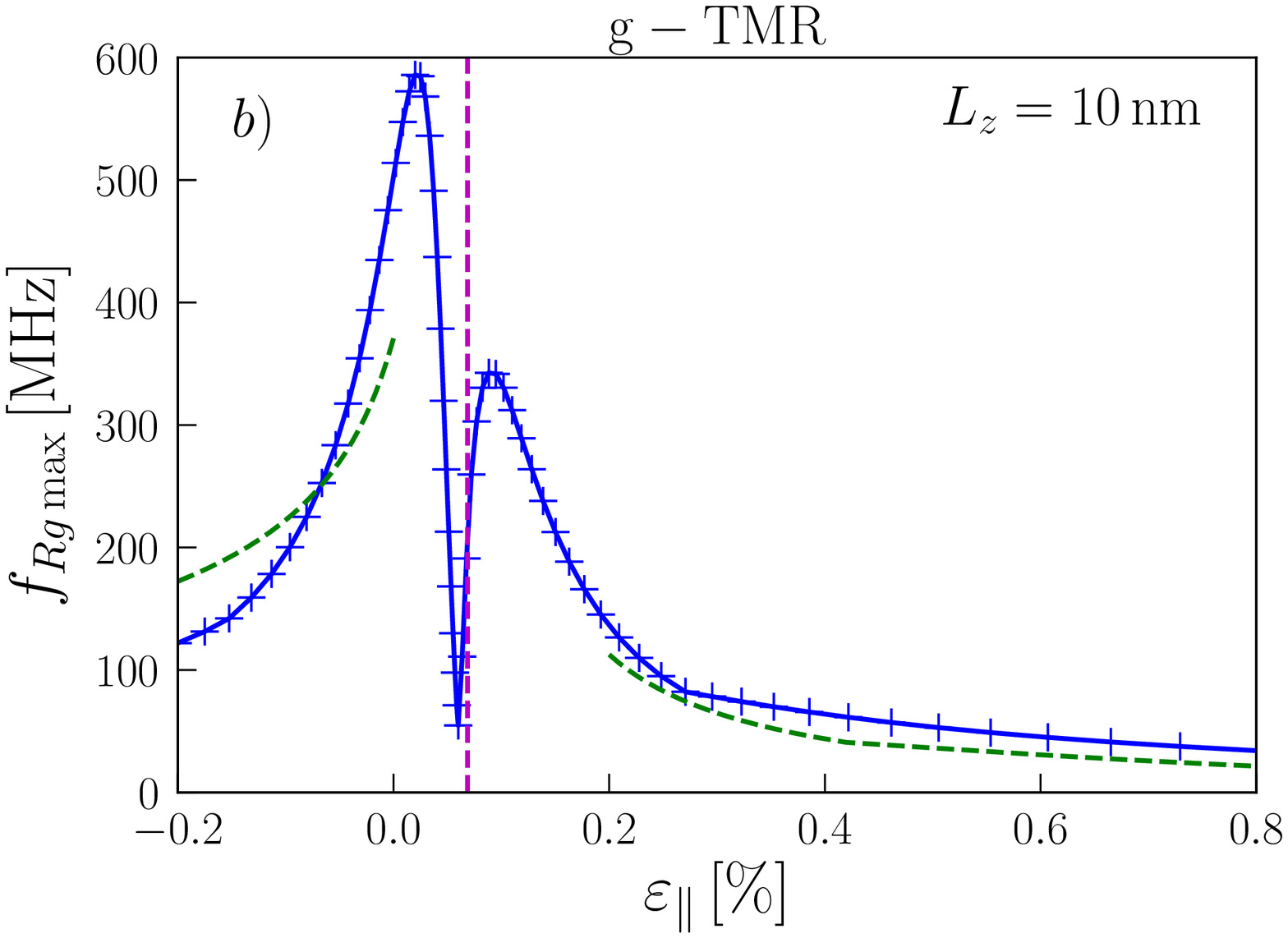}\\
  \includegraphics[width=0.4\textwidth]{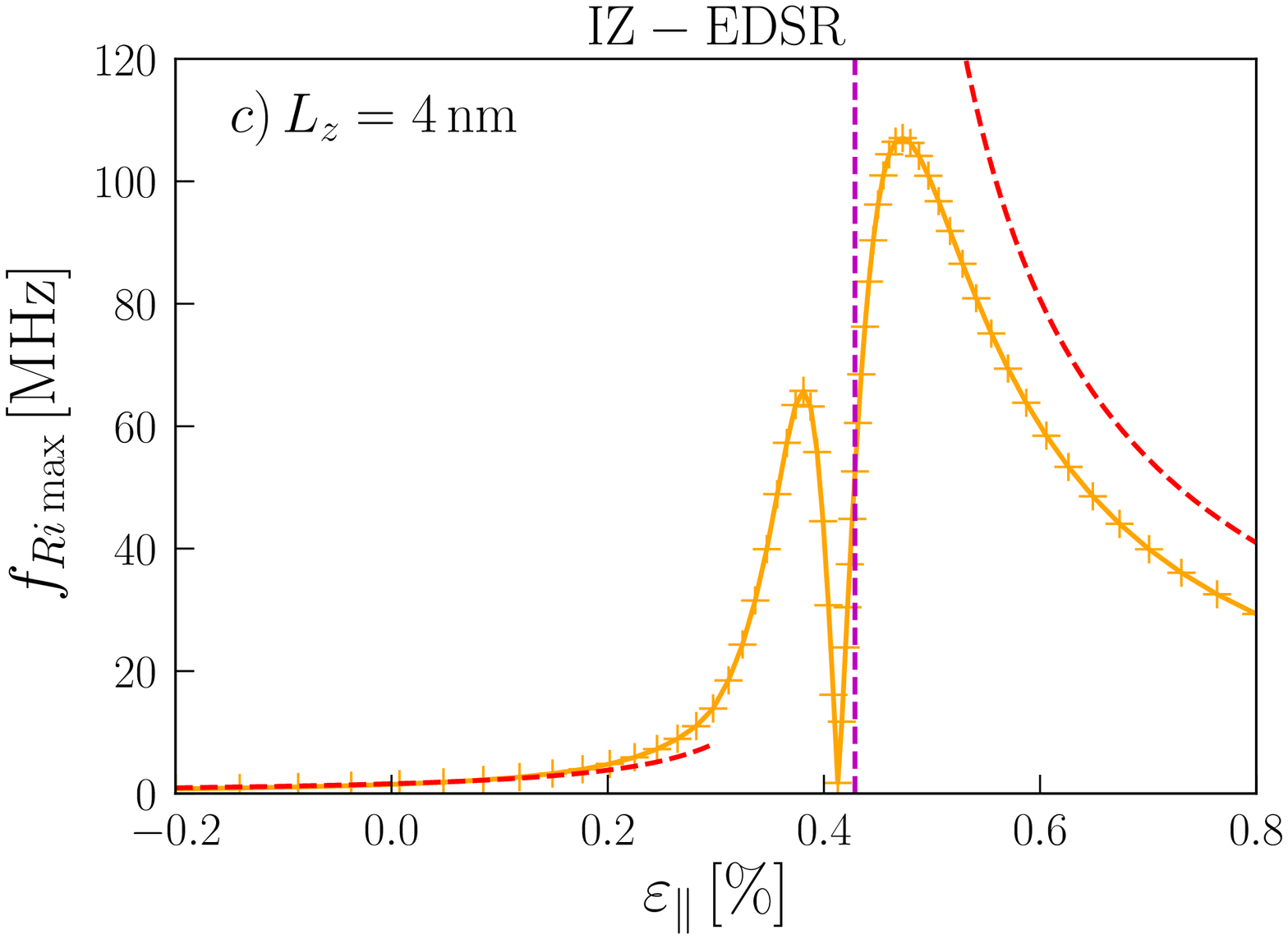}
  \includegraphics[width=0.4\textwidth]{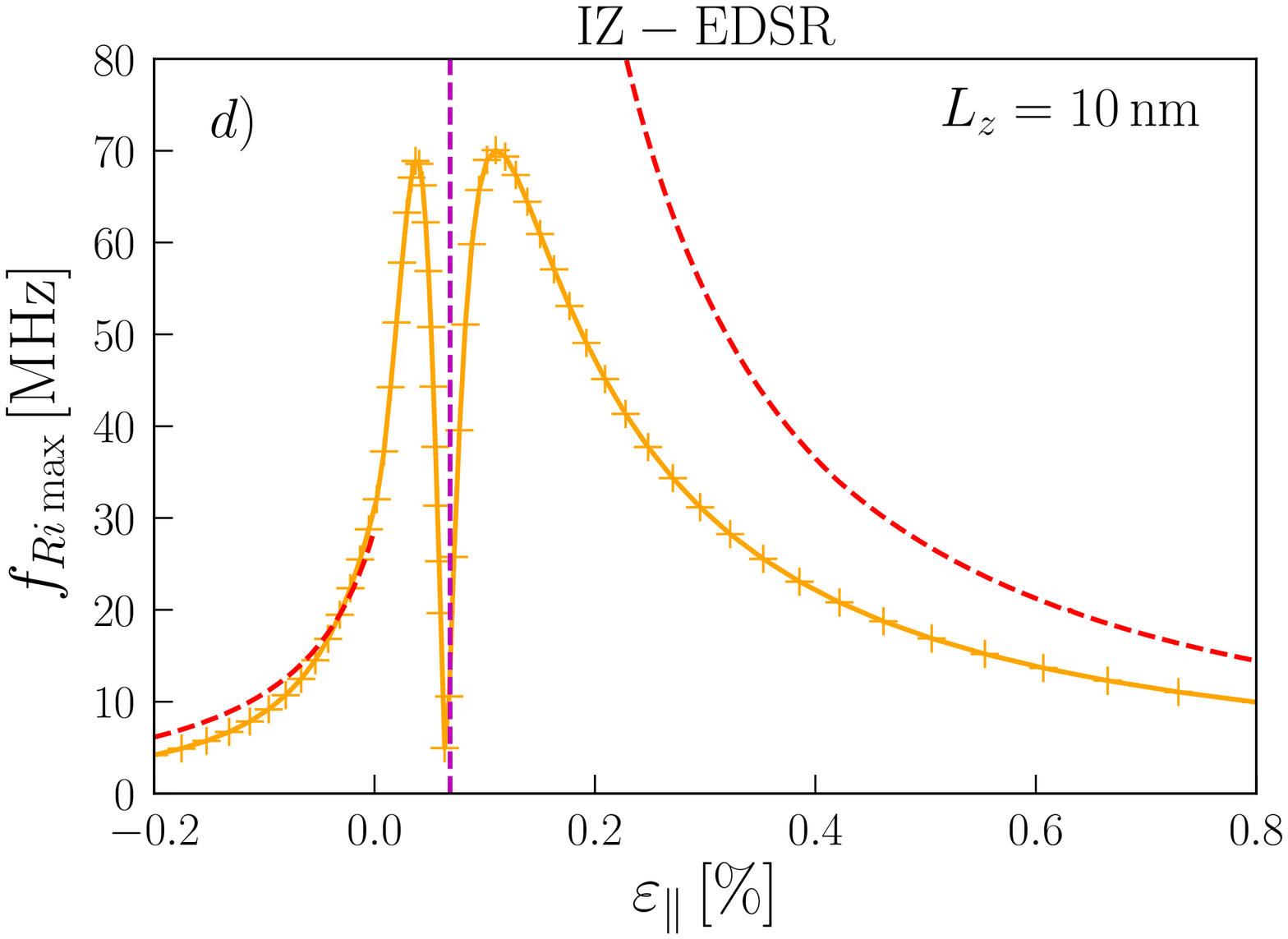}
\caption{$g$-TMR [a) and b)] and IZ-EDSR [c) and d)] Rabi frequencies dependence on biaxial strain. We compare the numerics\cite{Comment_numerics} (full line with symbols) with the semi-analytical formulas (dashed lines). The dashed vertical lines in magenta mark the critical strain $\varepsilon_\parallel^{\ast}$ that separates the heavy- and light-hole ground states. The material is silicon and the parameters are $L_z=4\unit{nm}$ for Figures a) and c), $L_z=10\unit{nm}$ for Figures b) and d), $B=1\unit{T}$, $\ef_{x/y}^{ac}=(1/30)\unit{mV/nm}$, $L_y=30\unit{nm}$, and $x_0=10\unit{nm}$.}
\label{fig:strain}
\end{figure*}

\section{Effect of biaxial strain, hole with mainly light character}\label{sec:strain}

The $g$-TMR and IZ-EDSR Rabi frequencies both depend on the gap $\Delta$ that rules the mixing between heavy- and light-hole states by lateral confinement and electric fields. Rabi oscillations are indeed forbidden in the absence of such mixing as discussed in Refs. \citenum{Kloeffel2018} and \citenum{Venitucci2019}. Biaxial strain in the ($xy$) plane controls the magnitude of this gap, and can even switch the character of the ground state \cite{Venitucci2019}. With strain the gap Eq. (\ref{Delta}) indeed becomes:
\begin{equation}\label{Delta_strain}
\Delta\rightarrow\Big|\frac{2\pi^2\gamma_2\hbar^2}{m_0L_z^2}-\Delta_{BP}\Big|.
\end{equation}
Here $\Delta_{BP}=-2(\nu+1)b_v\varepsilon_\parallel$ is the Bir-Pikus energy shift due to biaxial strain \cite{Venitucci2019,BirPikusBook}, with $\nu=2c_{12}/c_{11}$ the biaxial Poisson ratio, $c_{11}$, $c_{12}$ the elastic constants of the semiconductor, $b_v$ the uniaxial valence band deformation potential, and $\varepsilon_\parallel$ the in-plane strain. In the regime $\Delta_{BP}\gg 2\pi^2\gamma_2\hbar^2/(m_0L_z^2)$ the ground state has a dominant light-hole character. The transition from from a mostly heavy- to a mostly light-hole ground state actually takes place at strain:
\begin{equation}
\varepsilon_{\parallel}^{\ast}=\frac{\pi^2\gamma_2\hbar^2}{(\nu+1)|b_v|m_0L_z^2}.
\end{equation}
Note that $\varepsilon_{\parallel}^{\ast}$ can be very small in silicon ($\varepsilon_{\parallel}^{\ast}=0.069\%$ at $L_z=10$ nm), so that the transition to a light-hole ground state may possibly result from non-intentional process and cooldown strains \cite{Venitucci2018}.

With the methods of Section \ref{sec:HZ} we calculate the $g$-tensor corrections due to the coupling between the light-hole and the heavy-hole states. The $g$-tensor of a light-hole state is also diagonal and, by including the dominant perturbative corrections, its elements read
\begin{subequations}\label{g_l}
\begin{align}
 &g_x^{l}=-4\kappa+\delta g_{\parallel}^l+\delta g_{x}^l,\\
 &g_y^{l}=-4\kappa-\delta g_{\parallel}^l+\delta g_{y}^l,\\
 &g_z^{l}=-2\kappa-2\gamma_{l,1},
\end{align}
\end{subequations}
where $\delta g_{\parallel}$ is defined as
\begin{equation}\label{eq:g_para_l}
\delta g_\parallel^l=\frac{6\gamma_3\kappa\hbar^2}{m_0\Delta}\Big(\frac{\pi^2\f(\alpha_\parallel^l)}{L_y^2}-\frac{1}{2\lx^2}\Big),
\end{equation}
which is analogous to Eq. (\ref{eq:g_para_l}) but computed with the gap Eq. (\ref{Delta_strain}) and the in-plane light-hole effective mass
\begin{equation}\label{m_para_l}
m_\parallel^{l}=\frac{m_0}{\gamma_1-\gamma_2-\gamma_{l,1}}.
\end{equation}
The parameter $\gamma_{l,1}$ is
\begin{equation}\label{gammal}
 \gamma_{l,1}=\frac{6\gamma_3^2\hbar^2}{m_0}\sum_{n}\frac{|\langle\psi_1^l|k_z|\psi_n^h\rangle|^2}{E_1^l-E_n^h},
\end{equation}
where the energy denominator includes the Bir-Pikus energy shift. In an infinite square well potential along $z$ the matrix elements of the numerator of Eq. (\ref{gammal}) are the same as for heavy holes [Eq. (\ref{gammah})]. 

We compute the $g$-TMR Rabi frequency with the $g$-matrix formalism used in Sec. \ref{sec:gtmr}. We first neglect the corrections $\delta g_x^l$ and $\delta g_y^l$ that are due to the vector potential and are calculated in Appendix \ref{ap:g_corr}.
%\begin{equation}\label{fRgl}
% f_{Rg}^l=\frac{\mu_B\ef_{y}^{ac}\sqrt{(g_z^lB_zB_\parallel)^2+\big((g_x^l\frac{\partial g_{y}^l}{\partial\ef_y}-g_y^l\frac{\partial g_{x}^l}{\partial\ef_y})B_xB_y\big)^2}}{2h\sqrt{(g_x^lB_x)^2+(g_y^lB_y)^2+(g_z^lB_z)^2}}.
%\end{equation}
With the $g$-tensor elements given above we find that in strongly strained silicon and germanium the Rabi frequency is maximized for magnetic field components $(B_x,B_y,B_z)=(\pm\frac{B}{\sqrt{2}},\pm\frac{B}{\sqrt{2}},0)$ and reaches:
\begin{equation}\label{fRgmaxl}
 f_{Rg\max}^l\approx\frac{6\gamma_3|\kappa|\mu_B Be\ef_y^{ac}L_y\f'(\alpha_\parallel^l)}{\pi h(m_0/m_\parallel^l)\Delta}.
\end{equation}

We also note that the IZ-EDSR frequency for the light holes remains of the form of Eq. (\ref{fRipara}) with the corresponding effective mass and effective g-factors [Eqs. (\ref{m_para_l}) and (\ref{g_l})]. Then the maximal IZ-EDSR frequency is:
\begin{equation}\label{fRiparamax_l}
f_{Ri\parallel\max}^l=\frac{\delta x}{h\ell_{so\parallel}}\max(|g_x^l|,|g_y^l|)\mu_B B.
\end{equation}

In Fig. \ref{fig:map} we plot the color maps of the $g$-TMR and IZ-EDSR Rabi frequencies as functions of the parallel electric field $\ef_y$ and the in-plane strain. In Figs. (\ref{fig:comparison}) and (\ref{fig:strain}) we show the dependences of the Rabi frequencies on the electric field and strain respectively and we compare the semi-analytical formulas (that include the anisotropic corrections $\delta g_x$ and $\delta g_y$) with the numerical calculations. We discuss these figures in the next section. 

\section{Discussion}\label{sec:discussion}

\subsection{Comparison between $g$-TMR and IZ-EDSR}

The similarities between the static electric field configurations of Sections \ref{sec:gtmr} and \ref{sec:iso_para} allow for comparison between $g$-tensor magnetic resonance and iso-Zeeman electric dipole spin resonance effects. In the regime $\ell_{\ef_y}\gg L_y/\pi$ the ratio of the linear in static electric field Rabi frequencies is (neglecting the corrections of Appendix \ref{ap:g_corr})
\begin{equation}\label{eq:ratio}
\frac{f_{Rg\max}}{f_{Ri\parallel\max}}\approx\frac{|\kappa|}{\gamma_2|g_\parallel|}\frac{m_0}{m_\parallel}\frac{\ef_y^{ac}}{\ef_x^{ac}}\Big(\frac{L_y}{\pi\lx}\Big)^4.
\end{equation}  
With this it is clear that a key factor in the relative efficiency of $g$-TMR and IZ-EDSR manipulations is the ratio between the characteristic lengths of confinement along $x$ and $y$. If $\lx\gg L_y/\pi$ (with $\ef_x^{ac}$ and $\ef_y^{ac}$ of comparable magnitudes) then IZ-EDSR can be faster than $g$-TMR oscillations as illustrated in Fig. \ref{fig:Ey_fR_x015nm} of Appendix \ref{ap:add_figures}. For heavy holes however it turns out that $g_\parallel^h$ is quite small, which limits the IZ-EDSR Rabi frequencies ($g_\parallel^h\propto L_z^2$ so that $f_{Ri\parallel\max}^h\propto L_z^4$ when $L_z\to0$). On the contrary, we expect much more efficient IZ-EDSR manipulation for light holes (see Figs. \ref{fig:comparison} and \ref{fig:strain}).

Alternatively, the static electric field may be applied along $z$ in order to lift this limitation on the g-factor of heavy holes; yet polarizing the hole envelope in this direction is more challenging because of the strong confinement. In fact, the IZ-EDSR Rabi frequency $f_{Ri\perp\max}^h$ of heavy holes for static electric field $\ef_z$ perpendicular to the thin film remains of the same (fourth) order with respect to $L_z/\min(L_y, \pi\ell_{\ef_y})\ll 1$ (see Appendix \ref{ap:Hso_perp}). For $L_y/\pi<\ell_{\ef_y}$ we find\cite{Bulaev2007, Marcellina2017, Terrazos2020} that $f_{Ri\parallel\max}^h/f_{Ri\perp\max}^h\sim(\gamma_1/\gamma_2)(\ef_y/\ef_z)$ when $\gamma_2\ll\gamma_1$. 
In silicon, the IZ-EDSR shall therefore be much more efficient when the static electric field is parallel to the thin film (along $y$) than when it is perpendicular (along $z$). We have numerically verified (see Fig. \ref{fig:Eyz_fRi} of Appendix \ref{ap:add_figures}) that it is indeed the case when $\ef_y\sim\ef_z$. In germanium, however, the two configurations show Rabi frequencies with comparable magnitudes for electric fields in the few $\unit{mV/nm}$ range. For light holes on the other hand $f_{Ri\parallel\max}^l/f_{Ri\perp\max}^l\sim(\gamma_1/\gamma_3)(\ef_y/\ef_z)(\langle k_y^2\rangle L_z^2)^{-1}$, when the gap $\Delta$ including strain is of the order of the confinement gap at zero strain ($\varepsilon_\parallel\simeq 2\varepsilon_\parallel^\ast$). This is typically large since $\langle k_y^2\rangle L_z^2\ll 1$. Applying the static electric field along $y$, as done in this study, is therefore always much more efficient.

Fig. \ref{fig:map} represents the color maps of the numerically computed Rabi frequencies as a function of the static electric field and biaxial strain. Figures \ref{fig:map}a and \ref{fig:map}b show the $g$-TMR Rabi frequency for $L_z=4\unit{nm}$ and $L_z=10\unit{nm}$ while Figures \ref{fig:map}c) and \ref{fig:map}d) show the IZ-EDSR Rabi frequency for $L_z=4\unit{nm}$ and $L_z=10\unit{nm}$. 
The $g$-TMR and IZ-EDSR Rabi frequencies vanish along the line $\ef_y=0$. Breaking the inversion symmetry of the channel with a static electric field is indeed a pre-requisite for both mechanisms\cite{Kloeffel2018,Venitucci2018,Venitucci2019}. 
%At finite strain there are also regions where the driving can be strongly suppressed. For $g$-TMR they are at constant strain and were understood as lines where the the heavy and light hole envelopes are identical \cite{Venitucci2019}.

There is also a quasi-horizontal dip visible on Fig. \ref{fig:map}, near (but not at) the strain $\epsilon_\parallel^\ast$ where the heavy- and light-hole ground-states anti-cross ($\Delta=0$). This feature is also clearly visible on Fig. \ref{fig:strain}, and has already been identified in Ref. \citenum{Venitucci2019}. As it takes place near $\Delta = 0$, it is not captured by the present semi-analytical models. It actually arises when the qubit states and the excited states that are coupled by the static and $ac$ electric fields share very similar Bloch functions. Indeed, the Zeeman Hamiltonian can not couple such states (because their envelopes are, by design, orthogonal), so that the real space motion induced by the $ac$ electric field does not come along with pseudo-spin rotations. In general, this condition is met near $\Delta = 0$, because the qubit states rapidly switch from almost pure heavy- to almost pure light-hole states, and therefore cross the composition of the relevant excited states. 

In Figure \ref{fig:comparison} we show the Rabi frequencies computed semi-analytically with Eqs. (\ref{fRg}), (\ref{g_h_corr}), (\ref{g_l}), and (\ref{fRiparamax}) as a function of the static electric field $\ef_y$ and we compare them with the numerical calculations based on the four-band LK model. Figures \ref{fig:comparison}a and \ref{fig:comparison}b are computed at zero strain whereas Figures \ref{fig:comparison}c and \ref{fig:comparison}d are computed at $\varepsilon_\parallel = 0.7\%$ where the ground-state is mostly light-hole. In the thin film regime $L_z\ll L_y$ and for small electric fields such that $\ell_{\ef_y}>L_y/\pi$ we note a good correspondence between the analytics and the numerical calculations for both  $g$-TMR and IZ-EDSR. We also correctly predict the electric field optimum for the $g$-TMR Rabi frequency at $\ell_{\ef_y}\sim L_y/\pi$. For a stronger electric field such that $\ell_{\ef_y}<L_y/\pi$ the analytical expressions can significantly differ from the numerics. We attribute these discrepancies to deviations from the applicability of the lowest order of the perturbation theory, because the thin film condition may not be strictly fulfilled. 

We note that the $g$-TMR Rabi frequency shows an maximum at rather weak electric field while the IZ-EDSR Rabi frequency increases continuously over a wide range of electric field (see Fig. \ref{fig:lso}). The $g$-TMR Rabi frequency indeed decreases rapidly once the hole is squeezed by the static electric field $\ef_y$ and cannot be dragged efficiently anymore by the $ac$ electric field along $y$. On the contrary, the motion along $x$ is little hampered, and the direct Rashba spin-orbit coupling responsible for the IZ-EDSR oscillations is enhanced over a wide range of $\ef_y$. For heavy holes in silicon however, $g$-TMR remains more efficient than IZ-EDSR over the practical range of fields reached at low inversion density in CMOS devices typical of Refs. \citenum{Maurand2016} and \citenum{Crippa2018}. $g$-TMR shows, nonetheless, a more complex dependence on the magnetic field orientation (the optimal orientation showing, in particular, dot-to-dot variability, as suggested by Eq. (\ref{thetamax})). For light holes in silicon, IZ-EDSR can be more efficient than $g$-TMR at moderate electric fields. IZ-EDSR requires, on the other hand, at least two gates (for confinement and manipulation), whereas $g$-TMR can be achieved with one single gate both confining the hole and shaking the dot\cite{Venitucci2018}. As discussed above, an other way to promote IZ-EDSR over $g$-TMR (even for heavy holes) is to make the potential softer along $x$ ($\lx\gg L_y/\pi$) in order to enhance the motion of the dot, at the possible expense of an increased sensitivity to disorder along the channel.
	
Fig. \ref{fig:strain} shows the Rabi frequencies as a function of strain at a fixed electric field. The $g$-TMR and IZ-EDSR Rabi frequencies show a complex dependence on strain near $\varepsilon_\parallel=\varepsilon_\parallel^\ast$, characterized by a broad peak (due to the enhanced heavy- and light-hole mixing) split by the dip discussed above. The Rabi frequencies do decrease at large positive (tensile) and negative (compressive) strains as the heavy- and light-hole mixing gets inhibited by the increasing $|\Delta|$. 

The spin-orbit coupling strength in the devices can, therefore, be tuned by strains then further modulated by the static electric field. This does not only rule the Rabi frequency, but also the relaxation and coherence times of the qubits\cite{Li2020}. Compressive strains for example (as encountered in epitaxial germanium layers \cite{Lawrie2019,Scappucci2020,vanRiggelen2020,Hendrickx2020,Hendrickx2020_four_qubit}) do stabilize an almost pure heavy-hole, which can increase  $T_1$ and $T_2$ even faster than it decreases the Rabi frequency ($f_R$ being proportional to a dipole matrix element, and the electrical contributions to $1/T_1$ and $1/T_2^*$ to a dipole matrix element squared, except for quasi-static $1/f$ noise \cite{Paladino2014}). This might also ease the management and reduce variability in exchange interactions. Strains and electric fields must, therefore, be carefully engineered in order to optimize the overall performances of the devices.

\subsection{Dependence on material and channel orientation}

The Rabi frequencies depend on the host material through the Luttinger parameters $\gamma_1$, $\gamma_2$, $\gamma_3$, and through the Zeeman parameter $\kappa$. In order to compare channel materials and orientations, we have extracted the material-dependent prefactors of the $g$-TMR and IZ-EDSR Rabi frequencies of heavy ($h$) and light ($l$) holes. In the small electric field regime $\ell_{\ef_y}>L_y/\pi$, the maximal Rabi frequencies are proportional to:
\begin{subequations}\label{zeta_110}
\begin{align}
 &\zeta_{[110]}^{g{\rm -TMR},h}=\frac{\gamma_3\max(|\kappa-2\gamma_3\eta_{h,1}|,|\kappa-2\gamma_2\eta_{h,1}|)}{\gamma_2(\gamma_1+\gamma_2-\gamma_{h,1})^2},\\
 &\zeta_{[110]}^{{\rm IZ-EDSR},h}=\frac{\gamma_3^2\max(|\kappa-2\gamma_3\eta_{h,1}|,|\kappa-2\gamma_2\eta_{h,1}|)}{\gamma_2(\gamma_1+\gamma_2-\gamma_{h,1})^2},\\
 &\zeta_{[110]}^{g{\rm -TMR},l}=\frac{\gamma_3|\kappa+(\gamma_3-\gamma_2)\eta_{l,1}|}{(\gamma_1-\gamma_2-\gamma_{l,1})^2},\\
 &\zeta_{[110]}^{{\rm IZ-EDSR},l}=\frac{\gamma_2\gamma_3|\kappa|}{(\gamma_1-\gamma_2-\gamma_{l,1})^2}.
\end{align}
\end{subequations} 
The $g$-factor corrections calculated in Appendix \ref{ap:g_corr} ($\eta$ terms) are included in these prefactors. The $\zeta_{[110]}$'s of heavy holes are computed at zero strain, where the gap $\Delta$ is set by vertical confinement. Those of light-holes are computed at the same $\Delta$ that we assume controlled by strains. The values of the $\zeta_{[110]}$'s in silicon and germanium are collected in table \ref{tab:comparison}. We emphasize that the $\zeta_{[110]}$'s are intended for a comparison between different materials for a given mechanism, but not for a comparison between different mechanisms.

\begin{table}
\begin{tabular}{ c r r }
  \hline
  \hline
  \noalign{\smallskip}
  Material parameters & Si & Ge \\
  \noalign{\smallskip}
  \hline
  \noalign{\smallskip}
  $\gamma_1$ & 4.29  & 13.38 \\
  $\gamma_2$ & 0.34 & 4.24 \\
  $\gamma_3$ & 1.45 & 5.69 \\
  $\kappa$ & $-0.42$ & 3.41 \\
  $\gamma_{h,1}$ & 1.16 & 3.56 \\
  $\eta_{h,1}$ & 0.08 & 0.20\\
  $b_{v}\,[\unit{eV}]$ & $-2.10$ & $-2.86$\\
  $\nu=2c_{12}/c_{11}$ & 0.77 & 0.73\\
  \noalign{\smallskip}
  \hline
  \noalign{\smallskip}
  $\zeta_{[110]}^{g{\rm -TMR},h}$ & 0.23 & 0.012\\
  \noalign{\smallskip}
  \hline
  \noalign{\smallskip}
  $\zeta_{[110]}^{{\rm IZ-EDSR},h}$ & 0.34 & 0.067 \\
  \noalign{\smallskip}
  \hline
  \noalign{\smallskip}  
  $\zeta_{[110]}^{g{\rm -TMR},l}$ & 0.064 & 0.33\\
  \noalign{\smallskip}
  \hline
  \noalign{\smallskip}
  $\zeta_{[110]}^{{\rm IZ-EDSR},l}$& 0.013 & 0.98\\
  \noalign{\smallskip}
  \hline
  \hline
\end{tabular}
\caption{Rabi frequency dependence on the materials and comparison between silicon and germanium. The scaling factors $\zeta_{[110]}$ are those of Eqs. (\ref{zeta_110}). We evaluate the heavy-hole parameter $\gamma_{h,1}$ as well as $\zeta_{[110]}^{g{\rm -TMR},h}$ and $\zeta_{[110]}^{{\rm IZ-EDSR},h}$ at zero strain. For the light-hole case we take the large strain limit such that the energy gap between the light-hole and the heavy-hole states is dominated by the strain instead of the structural confinement, and the parameters $\gamma_{l,1}\ll\gamma_1-\gamma_2$ and $\eta_{l,1}\approx 1$. We emphasize that the numbers here illustrate the differences between silicon and germanium through their material dependent parameters; however they are not meant for a comparison between $g$-TMR and IZ-EDSR, nor for a comparison between the heavy-hole and light-hole cases. The Luttinger and strain parameters are borrowed from Ref. \citenum{Venitucci2019}.}
\label{tab:comparison}
\end{table}

In table \ref{tab:comparison} we note clear differences between silicon and germanium. As a main trend, electrically driving a heavy hole is expected to be more efficient in silicon than in germanium (for a given dot size). Indeed, the Rabi frequency at given static electric and magnetic fields is inversely proportional to a Luttinger parameter (IZ-EDSR) or to a Luttinger parameter squared ($g$-TMR), because heavier particles respond stronger to the static electric field $\ef_y$.\cite{Venitucci2019} Also, heavy holes benefit from the strong anisotropy of the valence band of silicon (large $\gamma_3/\gamma_2$ ratio). As a consequence of this anisotropy, the coupling between heavy and light holes by lateral confinement (driven by $\gamma_3$) is strong with respect to their splitting $\Delta$ ($\propto\gamma_2$), which enhances the heavy- and light-hole mixing in the qubit states and low-lying excitations, a pre-requisite for Rabi oscillations\cite{Kloeffel2018,Venitucci2019}. The advantage of silicon is even greater if the Rabi frequencies are compared at the same Zeeman splitting rather than the same magnetic field, as, in a first approximation, the $\zeta_{[110]}$'s must be rescaled by a factor $\simeq 1/\kappa$. On the contrary driving a light hole is expected to be more efficient in germanium, especially for IZ-EDSR that is almost two orders of magnitude stronger in germanium than in silicon. As a matter of fact, the gap $\Delta$ of light-hole qubits is primarily controlled by strains, so that silicon looses the benefits of its valence band anisotropy ($\gamma_2$ disappears from the denominator of the $\zeta_{[110]}$'s). Therefore, dealing with light holes in germanium may be interesting, but will require complex strain engineering. We would like, finally, to emphasize that the dots may be made larger in germanium than in silicon thanks to the lighter hole masses (reduced sensitivity to disorder), which can enhance the Rabi frequency of both heavy- and light-hole qubits. In particular, germanium hole qubits systematically perform better than silicon qubits if compared at the same vertical and lateral confinement energies (same $\gamma_2/L_z^2$, $(\gamma_1\pm\gamma_2-\gamma_{h/l,1})/L_y^2$ and $(\gamma_1\pm\gamma_2-\gamma_{h/l,1})/\lx^2$, which suppresses the demoninators of Eqs. (\ref{zeta_110})).

We also highlight the importance of the choice of the device orientation that expresses through the two parameters $\gamma_2$ and $\gamma_3$. Indeed, if the orientation of the channel ($x$ axis) is changed from $[110]$ to $[100]$ (and the $y$ axis from $[-110]$ to $[010]$), then $\gamma_2$ and $\gamma_3$ must be exchanged in the term $R$ [Eq. (\ref{R_Lutt})]:
\begin{equation}\label{R_Lutt_100}
R\rightarrow\frac{\hbar^2}{2m_0}\sqrt{3}[-\gamma_2(k_x^2-k_y^2)+2i\gamma_3k_xk_y],
\end{equation}
With this transformation the material-dependent prefactors become:
\begin{subequations}
\begin{align}
 &\zeta_{[100]}^{g{\rm -TMR},h}=\frac{\max(|\gamma_2(\kappa-2\gamma_3\eta_{h,1})|,|\gamma_2\kappa-2\gamma_3^2\eta_{h,1}|)}{\gamma_2(\gamma_1+\gamma_2-\gamma_{h,1})^2},\\
 &\zeta_{[100]}^{{\rm IZ-EDSR},h}=\frac{\gamma_3\max(|\gamma_2(\kappa-2\gamma_3\eta_{h,1})|,|\gamma_2\kappa-2\gamma_3^2\eta_{h,1}|)}{\gamma_2(\gamma_1+\gamma_2-\gamma_{h,1})^2},\\
 &\zeta_{[100]}^{g{\rm -TMR},l}=\frac{|\gamma_2\kappa+\gamma_3(\gamma_2-\gamma_3)\eta_{l,1}|}{(\gamma_1-\gamma_2-\gamma_{l,1})^2},\\
 &\zeta_{[100]}^{{\rm IZ-EDSR},l}=\frac{\gamma_2\gamma_3|\kappa|}{(\gamma_1-\gamma_2-\gamma_{l,1})^2}.
\end{align}
\end{subequations} 
We give in Table \ref{tab:comparison_100} the values of the $\zeta_{[100]}$'s for silicon and germanium. For a heavy hole the $\zeta_{[100]}$'s are smaller than the $\zeta_{[110]}$'s for both $g$-TMR and IZ-EDSR, so that the $[110]$ orientation is optimal in this case. This largely results for silicon from the loss of the $\sim\gamma_3/\gamma_2$ enhancement factor related to the valence band anisotropy\cite{Venitucci2019} (both the gap $\Delta$ and the coupling between heavy and light holes being ruled by $\gamma_2$ in the $[100]$ orientation). For the $g$-TMR of a light hole $\zeta_{[100]}^{g{\rm -TMR},l}>\zeta_{[110]}^{g{\rm -TMR},l}$ for silicon and $\zeta_{[100]}^{g{\rm -TMR},l}<\zeta_{[110]}^{g{\rm -TMR},l}$ for germanium. For the IZ-EDSR of a light hole the prefactors are essentially the same  for the two orientations.

Therefore regarding silicon in the present configuration the choice of the $[110]$ orientation is optimal, at least for a heavy hole.

\begin{table}
\begin{tabular}{ c r r }
  \hline
  \hline
  \noalign{\smallskip}
  Material parameters & Si & Ge \\
  \noalign{\smallskip}
  \hline
  \noalign{\smallskip}
  $\zeta_{[100]}^{g{\rm -TMR},h}$ & 0.12 & 0.0058\\
  \noalign{\smallskip}
  \hline
  \noalign{\smallskip}
  $\zeta_{[100]}^{{\rm IZ-EDSR},h}$ & 0.17 & 0.033 \\
  \noalign{\smallskip}
  \hline
  \noalign{\smallskip}  
  $\zeta_{[100]}^{g{\rm -TMR},l}$ & 0.11 & 0.074\\
  \noalign{\smallskip}
  \hline
  \noalign{\smallskip}
  $\zeta_{[100]}^{{\rm IZ-EDSR},l}$& 0.013 & 0.98\\
  \noalign{\smallskip}
  \hline
  \hline
\end{tabular}
\caption{Rabi frequency material-dependent prefactors and comparison between silicon and germanium for a channel oriented along $[100]$.}
\label{tab:comparison_100}
\end{table}

\section{Conclusion}

We have examined the electrical manipulation of hole qubits in a 1D channels that resemble the MOS setup of Refs. \citenum{Maurand2016} and \citenum{Crippa2018} where the structural confinement is strong in one direction ($z$) and the most relevant static electric field is perpendicular to that direction ($y$). This configuration allows for stronger electrical polarizability than a static electric field along $z$.
We have compared two mechanisms of electrical manipulation, the $g$-tensor magnetic resonance ($g$-TMR, $ac$ electric field also parallel to $y$), and the iso-Zeeman electric dipole spin resonance (IZ-EDSR, $ac$ electric field along the channel direction $x$), and we have evaluated their efficiencies as given by the magnitudes of the Rabi frequencies. 
In the regime of weak mixing between the heavy-hole and the light-hole states we thoroughly analyze the spin-orbit interactions responsible for the two effects. In particular we derive the effective Rashba Hamiltonian, Eq. (\ref{HR1D_para}), that leads to the IZ-EDSR effect with Rabi frequency given by Eq. (\ref{fRipara}).
The two mechanisms can be controlled by the electric bias and by the strains, as highlighted in Fig. \ref{fig:map}. The $g$-TMR Rabi frequency is maximal at only moderate electric fields [Eq. (\ref{fRgmaxs})] while IZ-EDSR is optimal at stronger electric fields such that the energy of electrical confinement (along $y$) is comparable to the energy of the strong structural confinement (along $z$). For such strong electric fields IZ-EDSR can be the most efficient mechanism as shown in Fig. (\ref{fig:comparison}). In addition, the IZ-EDSR Rabi frequency strongly depends on the extent of the envelope function along the driving $ac$ field ($x$), as shown by Eq. (\ref{eq:ratio}).
Furthermore, we have discussed the effect of strains, which can notably switch the dominant character of the hole (Sec. \ref{sec:strain}). Moving from a heavy-hole to a light-hole qubit actually strengthens the IZ-EDSR owing to the dependence of the Rabi frequency on the in-plane $g$-factors in Eq. (\ref{fRiparamax}). The behavior of the Rabi frequencies with biaxial strain is illustrated in Fig. \ref{fig:strain}, which highlights particular values of $\varepsilon_\parallel$ near the heavy- to light-hole transition where the frequencies essentially vanish. The Rabi frequencies do also decrease at large compressive and tensile strains because of the reduce heavy- and light-hole mixing; this may however strongly increase lifetimes and reduce variability. Strains and electric fields must, therefore, be carefully engineered in order to optimize the overall performances of the qubits. Then we have discussed the choice of the host material and we have compared, in particular, electrical manipulation in silicon and in germanium. According to Table \ref{tab:comparison}, both $g$-TMR and IZ-EDSR are more efficient in silicon than in germanium quantum dots with the same size, in the weak electric field regime and in the absence of strains, due to the larger hole effective masses. However, when the qubit acquires a dominant light-hole character under tensile strain, germanium can be more efficient than silicon, especially in the IZ-EDSR configuration. Moreover, germanium systematically outperforms silicon if the dots are compared at different sizes but same vertical and lateral confinement energies. Tables \ref{tab:comparison} and \ref{tab:comparison_100} also show the influence of the cristallographic orientation of the channel. We find that for heavy holes the $[110]$ orientation is optimal as it takes best advantage of the anisotropy of the valence band of silicon. These conclusions provide guidelines for the design and optimization of hole spin-orbit qubits embedded in one-dimensional channels.

\section*{Acknowledgements}

This  work  was  supported  by  the  European  Union Horizon  2020  research  and  innovation  program  under grant  agreement 810504-QUCUBE-ERC-2018-SyG,  and  by  the  French  national research agency (ANR project MAQSi).

\appendix

\section{Effective Hamiltonians}\label{ap:EH}

We compute the Hamiltonians (\ref{Heff}) and (\ref{HR1D_para}) perturbatively with the effective Hamiltonian method as presented in Ref. \citenum{Atom-photon} for instance. The approach is justified when the eigenstates of the unperturbed Hamiltonian can be sorted into two groups $\alpha$ and $\beta$ with well separated energies. The effective Hamiltonian then describes the dynamics of the states of group $\alpha$ and gathers corrections due to the coupling with the states of group $\beta$. The second-order term of the perturbation series for the effective Hamiltonian is
\begin{eqnarray}
\nonumber \langle i\alpha|H_{\rm{eff},\alpha}^{(2)}|j\alpha\rangle=\frac{1}{2}\sum_{k,\beta\neq\alpha}\langle i\alpha|H'|k\beta\rangle\langle k\beta|H'|j\alpha\rangle\\
\times\Big(\frac{1}{E_i^\alpha-E_k^\beta}+\frac{1}{E_j^\alpha-E_k^\beta}\Big),
\end{eqnarray}
where $H'$ represents the coupling between the states of $\alpha$ and those of $\beta$.
Here, $\alpha$ are the qubit states (the topmost Kramers pair at zero magnetic field) and $\beta$ collects all the excited states.

Here we compute the effective Hamiltonians for qubit states that can be of heavy- or light-hole type. 
We take as a starting point the four-band LK Hamiltonian defined in Ref. \citenum{Venitucci2019}, and we neglect the split-off states because we assume that the spin-orbit gap is always larger than the gap due to the confinement [Eq. (\ref{Delta})]. 
With the notation of e.g. Ref. \citenum{Venitucci2019} the corrections at leading order in $L_z/\min(\ell_x,L_y/\pi,\ell_{\ef_y})$ appear at second order in the term
\begin{equation}
 S=\frac{\sqrt{3}\hbar^2}{m_0}\gamma_3(k_x-ik_y)k_z,
\end{equation}
and renormalize the effective mass Eq. (\ref{m_para_h}) and the transverse $g$-factor Eq. (\ref{gzh}) (Ref. \citenum{Ares2013}). Furthermore the Hamiltonian (\ref{Heff}) includes the contributions from the term $R$ and from the off-diagonal elements of the Zeeman Hamiltonian (D3) in Ref. \citenum{Venitucci2019}. The corrections to the $g$-factors that come from the electromagnetic vector potential (Appendix \ref{ap:g_corr}) include the cross terms of $R$ and $S$.
The Rashba spin-orbit Hamiltonian (\ref{HR1D_para}) derived in appendix \ref{ap:Hso_para} collects the cross terms given by the first and the second parts of $R$. The spin-orbit coupling of Appendix (\ref{ap:Hso_perp}) includes the cross terms of $R$ and $S$. 

\section{Corrections to the $g$-tensor Eq. (\ref{geff_para}) due to the electromagnetic vector potential}\label{ap:g_corr}

We calculate a correction to Eq. (\ref{geff_para}) that arises from the vector potential in the perturbation theory. In the thin film regime the canonical momenta couple the in-plane components of the magnetic field with the orbit of the hole\cite{Comment_gauge}: 
\begin{subequations}
\begin{align}
 &\tilde{k}_x=k_x+\frac{eB_y}{\hbar}z,\\
 &\tilde{k}_y=k_y-\frac{eB_x}{\hbar}z.
\end{align}
\end{subequations}
These terms are also known to give rise to an anisotropy in the in-plane effective $g$-factor for quasi-2D electrons in the presence of Dresselhaus spin-orbit coupling, see Refs. \citenum{Kalevich1993,Falko2005,Michal2018,Stano2018}.
The effective magnetic Hamiltonian collects the cross terms of $R$ and $S$ at second order in the coupling between the heavy-hole and the light-hole states: 
\begin{widetext}
\begin{equation}
H_{{\rm eff},h}^{(2)}=\sum_{n}\frac{\langle\psi_1^h|R|\psi_n^l\rangle\langle\psi_n^l|S|\psi_1^h\rangle-\langle\psi_1^h|S|\psi_n^l\rangle\langle\psi_n^l|R|\psi_1^h\rangle}{E_1^h-E_n^l}\sigma_++H.c.
\end{equation}
\end{widetext}

In the above effective Hamiltonian the vector potential has no influence on the term $S$ at first order in the magnetic field because of the cancellation of the symmetrized product: $\langle\psi_1^h|zk_z+k_zz|\psi_1^h\rangle=0$. It however enters in the term $R$ through the products: 
\begin{subequations}
\begin{align}
 &\tilde{k}_x^2=k_x^2+\frac{2eB_y}{\hbar}zk_x+O(B_y^2),\\
 &\tilde{k}_y^2=k_y^2-\frac{2eB_x}{\hbar}zk_y+O(B_x^2),\\
 &\tilde{k}_x\tilde{k}_y=k_xk_y+\frac{eB_y}{\hbar}zk_y-\frac{eB_x}{\hbar}zk_x+O(B_xB_y).
\end{align}
\end{subequations}
Collecting the terms linear in the magnetic field and, considering as before that $\langle k_xk_y\rangle=0$, we obtain the following correction to the Zeeman Hamiltonian of heavy holes:
\begin{eqnarray}\label{correction_Heff}
\nonumber \delta H_{Z}^{h}&=&\frac{6\gamma_3\eta_{h,1}\hbar^2\mu_B}{m_0\Delta}[(\gamma_2\langle k_x^2\rangle-\gamma_3 \langle k_y^2\rangle)B_x\sigma_x\\
&&+(\gamma_3\langle k_x^2\rangle-\gamma_2\langle k_y^2\rangle)B_y\sigma_y],
\end{eqnarray}
where $\Delta$ is given by Eq. (\ref{Delta_strain}). We define the dimensionless parameter
\begin{eqnarray}\label{eta}
\eta_{h,1}=\Delta\sum_{n>1}\frac{2{\rm Im}(\langle\psi_1^h|z|\psi_n^l\rangle\langle\psi_n^l|k_z|\psi_1^h\rangle)}{E_1^h-E_n^l},
\end{eqnarray}
where, for $n>1$,
\begin{equation}\
{\rm Im}(\langle\psi_1^h|z|\psi_n^l\rangle\langle\psi_n^l|k_z|\psi_1^h\rangle)=\frac{8n^2((-1)^{n+1}-1)^2}{\pi^2(n^2-1)^3},
\end{equation}
and the energy separations between the states of the dot are:
\begin{equation}
E_1^h-E_n^l=-\Delta_{BP}+\frac{\pi^2\hbar^2}{m_0L_z^2}\big(\gamma_2(n^2+1)+\frac{\gamma_1}{2}(n^2-1)\big).
\end{equation}
With Eq. (\ref{correction_Heff}) we get the corrections to the diagonal elements of the $g$-tensor:
\begin{subequations}
\begin{align}\label{delta_g_h}
 &\delta g_{x}^h=\frac{12\gamma_3\eta_{h,1}\hbar^2}{m_0\Delta}\big(\gamma_2\langle k_x^2\rangle-\gamma_3\langle k_y^2\rangle\big),\\
 &\delta g_{y}^h=\frac{12\gamma_3\eta_{h,1}\hbar^2}{m_0\Delta}\big(\gamma_3\langle k_x^2\rangle-\gamma_2\langle k_y^2\rangle\big).
\end{align}
\end{subequations}
Together with Eq. (\ref{geff_para}) the corrected $g$-factors write:
\begin{subequations}
\begin{align}\label{g_h_corr}
 &g_{x}^h=\frac{6\gamma_3\hbar^2}{m_0\Delta}\big((\kappa-2\gamma_3\eta_{h,1})\langle k_y^2\rangle-(\kappa-2\gamma_2\eta_{h,1})\langle k_x^2\rangle\big),\\
 &g_{y}^h=\frac{6\gamma_3\hbar^2}{m_0\Delta}\big((\kappa-2\gamma_2\eta_{h,1})\langle k_y^2\rangle-(\kappa-2\gamma_3\eta_{h,1})\langle k_x^2\rangle\big).
\end{align}
\end{subequations}

These corrections break the rotational symmetry of the  $g$-tensor in the $(xy)$ plane. The new terms feature the dimensionless parameter $\eta_{h,1}$ that in the absence of strain is $\eta_{h,1}\approx0.08$ for silicon and $\eta_{h,1}\approx0.20$ for germanium, given the Luttinger parameters of Table \ref{tab:comparison}. 
Thus the corrections (\ref{delta_g_h}) have negative signs for a confinement much stronger in the $y$ direction than in the $x$ direction because $\gamma_2$ and $\gamma_3$ are positive for both silicon and germanium. Comparing with Eq. (\ref{geff_para}) we see that for silicon, whose parameter $\kappa$ is negative, the correction (\ref{delta_g_h}) enhances the Rabi frequency. On the contrary for germanium, which has positive $\kappa$, the correction reduces the Rabi frequency. These trends have been already observed in previous numerical calculations\cite{Venitucci2019}. For silicon and germanium the correction (\ref{delta_g_h}) changes the $g$-factors and the Rabi frequencies by a factor of order $1$. In the calculation of the silicon and germanium Rabi frequency in Figs. \ref{fig:comparison} and \ref{fig:strain} and in Appendix \ref{ap:add_figures} we have included this correction and verified the improved agreement between the analytics and the numerics in the thin film regime.

On the other hand for the light-hole ground state we have:
\begin{widetext}
\begin{equation}
H_{{\rm eff},l}^{(2)}=\sum_{n}\frac{\langle\psi_1^l|R|\psi_n^h\rangle\langle\psi_n^h|S^*|\psi_1^l\rangle-\langle\psi_1^l|S^*|\psi_n^h\rangle\langle\psi_n^h|R|\psi_1^l\rangle}{E_1^l-E_n^h}\sigma_++H.c.
\end{equation}
\end{widetext}
Taking the same steps as for the heavy-hole case we define the dimensionless parameter
\begin{eqnarray}\label{eta_l}
\eta_{l,1}=\Delta\sum_{n>1}\frac{2{\rm Im}(\langle\psi_1^l|z|\psi_n^h\rangle\langle\psi_n^h|k_z|\psi_1^l\rangle)}{E_1^l-E_n^h},
\end{eqnarray}
where $\Delta$ is given by Eq. (\ref{Delta_strain}), and we get the corrections:
\begin{subequations}
\begin{align}\label{delta_g_l}
 &\delta g_{x}^l=\frac{12\gamma_3\eta_{l,1}\hbar^2}{m_0\Delta}\big(\gamma_2\langle k_x^2\rangle+\gamma_3\langle k_y^2\rangle\big),\\
 &\delta g_{y}^l=\frac{12\gamma_3\eta_{l,1}\hbar^2}{m_0\Delta}\big(\gamma_3\langle k_x^2\rangle+\gamma_2\langle k_y^2\rangle\big).
\end{align}
\end{subequations}
Thus at leading order of perturbation the in-plane light-hole $g$-factors read (see Eq. (\ref{g_l})):
\begin{subequations}
\begin{align}\label{g_l_corr}
 &g_{x}^l=-4\kappa\nonumber+\frac{6\gamma_3\hbar^2}{m_0\Delta}\big((\kappa+2\gamma_3\eta_{l,1})\langle k_y^2\rangle-(\kappa-2\gamma_2\eta_{l,1})\langle k_x^2\rangle\big),\\
 &g_{y}^l=-4\kappa+\frac{6\gamma_3\hbar^2}{m_0\Delta}\big((-\kappa+2\gamma_2\eta_{l,1})\langle k_y^2\rangle+(\kappa+2\gamma_3\eta_{l,1})\langle k_x^2\rangle\big).
\end{align}
\end{subequations}
These corrections have also been included in Figs. \ref{fig:comparison} and \ref{fig:strain} and in Appendix \ref{ap:add_figures}.

\section{Derivation of the Rashba spin-orbit Hamiltonian Eq. (\ref{HR1D_para})}\label{ap:Hso_para}

We derive the effective Rashba spin-orbit Hamiltonian Eq. (\ref{HR1D_para}) in the simple thin film limit with an electric field oriented in the plane of the film. With the effective Hamiltonian method described in Appendix \ref{ap:EH}, the cross terms of $R$ yield:
\begin{widetext}
\begin{equation}\label{Heffso_full}
H_{so\parallel}=\frac{-3i\hbar^4\gamma_2\gamma_3}{2m_0^2}\sum_n\frac{\langle \chi_1^\alpha|k_y^2|\chi_n^\beta\rangle\langle \chi_n^\beta|k_y|\chi_1^\alpha\rangle-\langle \chi_1^\alpha|k_y|\chi_n^\beta\rangle\langle \chi_n^\beta|k_y^2|\chi_1^\alpha\rangle}{\Delta+E_{y,1}^\alpha-E_{y,n}^\beta}k_x\sigma_z.
\end{equation}
\end{widetext}
Here $\chi_n^{\alpha/\beta}$ are the envelope functions describing the motion of heavy and light holes along $y$ (the indices $\alpha$ and $\beta$ are unspecified and they can represent either light-hole or heavy-hole states), and we introduce the Pauli matrix $\sigma_z$ because the sign of the matrix elements depends on the pseudo-spin state of the hole. It is positive for $J_z=3/2$ and negative for $J_z=-3/2$.
We can simplify Eq. (\ref{Heffso_full}) very much in the small electric field limit. Indeed in this limit the sum in Eq. (\ref{Heffso_full}) converges rapidly and the energy denominator weakly depends on the indices of the relevant excited states, so that we approximate: 
\begin{eqnarray}\label{Heffso}
H_{so\parallel}&=&\frac{-3i\hbar^4\gamma_2\gamma_3}{2m_0^2\Delta}\sum_n\Big(\langle \chi_1^\alpha|k_y^2|\chi_n^\beta\rangle\langle \chi_n^\beta|k_y|\chi_1^\alpha\rangle\\
&&-\langle \chi_1^\alpha|k_y|\chi_n^\beta\rangle\langle \chi_n^\beta|k_y^2|\chi_1^\alpha\rangle\Big)k_x\sigma_z.
\end{eqnarray} 
Let us note that on its domain of support the envelope function of the hole satisfies 
\begin{equation}\label{Schroedinger_y}
\Big(-\frac{\hbar^2k_y^2}{2m_\parallel^\alpha}+e\ef_y y\Big)|\chi_1^\alpha\rangle=E_{y,1}^\alpha|\chi_1^\alpha\rangle.
\end{equation}
We substitute Eq. (\ref{Schroedinger_y}) in Eq. (\ref{Heffso}) and then we remove the sum over the intermediate states, since they constitute a complete basis in the subspace of the light-hole envelopes. We get:
\begin{equation}\label{H_so_linear}
H_{so\parallel}=\frac{-3i\hbar^2\gamma_2\gamma_3 m_\parallel^\alpha e\ef_y}{m_0^2\Delta}\langle \chi_1^\alpha|(yk_y-k_y y)|\chi_1^\alpha\rangle k_x\sigma_z.
\end{equation}
With the commutation relation $[y,k_y]=yk_y-k_y y=i$ we arrive at Eq. (\ref{HR1D_para}) with the inverse spin-orbit length defined by Eq. (\ref{l_so_inv_lin}).

Beyond the linear in electric field approximation given by Eq. (\ref{H_so_linear}) the inverse spin-orbit length that follows from Eqs. (\ref{Heffso_full}) and (\ref{HR1D_para}) reads
\begin{widetext}
\begin{equation}\label{l_so_inv}
\ell_{so\parallel}^{-1}=\frac{-3i\hbar^2\gamma_2\gamma_3 m_\parallel^\alpha }{2m_0^2}\sum_n\frac{\langle \chi_1^\alpha|k_y^2|\chi_n^\beta\rangle\langle \chi_n^\beta|k_y|\chi_1^\alpha\rangle-\langle \chi_1^\alpha|k_y|\chi_n^\beta\rangle\langle \chi_n^\beta|k_y^2|\chi_1^\alpha\rangle}{\Delta+E_{y,1}^\alpha-E_{y,n}^\beta}.
\end{equation}
\end{widetext}
In Fig. (\ref{fig:lso}) we compare Eqs. (\ref{l_so_inv}) and (\ref{l_so_inv_lin}) for the inverse spin-orbit length, and we use Eq. (\ref{l_so_inv}) in the calculation of the IZ-EDSR Rabi frequency in Figs. (\ref{fig:comparison}) and (\ref{fig:strain}).

\section{Hole IZ-EDSR in one dimension}\label{ap:EDSR}

We transform the IZ-EDSR Hamiltonian with spin-orbit coupling Eq. (\ref{HR1D_para}) and obtain the time-dependent Zeeman coupling Eq. (\ref{HZt}). Essentially we do the calculation of Ref. \citenum{Golovach2006} in the special one-dimensional case. For holes the EDSR Hamiltonian reads:
\begin{eqnarray}
\nonumber H & = & -\frac{\hbar^2k_x^2}{2m_\parallel}-\frac{1}{2}Kx^2+e\ef_x^{ac}(t)x\\
&&+\frac{\mu_B}{2}(g{\bf B})_a\sigma_a+\frac{\hbar^2}{m_\parallel\ell_{so\parallel}}k_x\sigma_z.
\end{eqnarray}
In the Zeeman term the summation over the repeated index $a$ is implicit.
We make the unitary transformation\cite{Golovach2006} $\tilde{H}=e^{O}He^{-O}$ that cancels the spin-orbit coupling term to first order in $\ell_{so\parallel}^{-1}$. We directly check that the operator
\begin{equation}
 O=-i\frac{x}{\ell_{so\parallel}}\sigma_z+i\frac{\mu_B}{K\ell_{so\parallel}}k_x\epsilon_{abz}(g{\bf B})_a\sigma_b
\end{equation}
does this, where $\epsilon$ is the Levi-Civita antisymmetric symbol. As a result,
\begin{eqnarray}
\nonumber \tilde{H} & = & -\frac{\hbar^2k_x^2}{2m_\parallel}-\frac{1}{2}Kx^2+e\ef_x^{ac}(t)x\\
&&+\frac{\mu_B}{2}(g{\bf B})_a\sigma_a+\delta H_Z(t),
\end{eqnarray}
with:
\begin{eqnarray}
\nonumber\delta H_Z(t)&=&[O,e\ef_x^{ac}(t)x]\\
\nonumber &=&\frac{e\ef_{x}^{ac}(t)}{K\ell_{so\parallel}}\mu_B\epsilon_{abz}(g{\bf B})_a\sigma_b\\
&=&\frac{\delta x(t)}{\ell_{so\parallel}}\mu_B[(g{\bf B})_x\sigma_y-(g{\bf B})_y\sigma_x].
\end{eqnarray}
If the $g$-tensor is diagonal this equals (\ref{HZt}). 

\section{IZ-EDSR in a perpendicular static electric field}\label{ap:Hso_perp}

Let us now consider the IZ-EDSR effect for a static electric field $\ef_z$ in the direction of the strong confinement. By perturbation of the LK model (appendix \ref{ap:EH}) we obtain the quasi-2D effective Rashba Hamiltonian \cite{Marcellina2017}: 
\begin{eqnarray}\label{Hso_perp}
H_{so\bot}^{(2D)}&=&A[\gamma_3(k_y^3\sigma_x+k_x^3\sigma_y)\nonumber\\
&-&(\gamma_3+2\gamma_2)(k_xk_yk_x\sigma_x+k_yk_xk_y\sigma_y)].
\end{eqnarray}
The prefactor of the above equation is
\begin{equation}\label{socp}
A=\frac{3\hbar^4\gamma_3}{m_0^2}\sum_n\frac{{\rm Im}(\langle\psi_1^h|\psi_n^l\rangle\langle\psi_n^l|k_{z}|\psi_1^h\rangle)}{E_1^h-E_n^l},
\end{equation}
which evaluates as $A\approx3\hbar^4\gamma_3(\alpha_\bot^h-\alpha_\bot^l)/(2m_0^2L_z\Delta)$, with
\begin{equation}\label{alpha_perp}
\alpha_\bot^{h/l}=\frac{2m_\bot^{h/l}e\ef_z L_z^3}{\hbar^2\pi^3}\ll1,
\end{equation}
where $m_\bot^{h/l}=m_0/(\gamma_1\mp2\gamma_2)$ are the vertical confinement masses of heavy and light holes, respectively.

When the parabolic confinement along $x$ is weak compared with the confinement along $y$, Eq. (\ref{Hso_perp}) reduces to quasi-1D Rashba Hamiltonian:
\begin{equation}\label{HR1D_perp}
 H_{so\bot}=-A(\gamma_3+2\gamma_2)\langle k_y^2\rangle k_x\sigma_y.
\end{equation}
This yields the inverse spin-orbit length of a heavy hole
\begin{eqnarray}
\ell_{so\bot}^{-1}&=&\frac{(\gamma_3+2\gamma_2)|A|\langle k_y^2\rangle m_\parallel^h}{\hbar^2}\nonumber\\
&\approx& \frac{3\gamma_3(\gamma_3+2\gamma_2)}{4\gamma_2(m_0/m_\parallel^h)}(\alpha_\bot^h-\alpha_\bot^l)\langle k_y^2\rangle L_z,
\end{eqnarray}
then the Rabi frequency:
\begin{equation}\label{fRi_perp}
f_{Ri\bot}^h=\frac{\delta x}{h\ell_{so\bot}}\mu_B\sqrt{(g_z^h B_z)^2+(g_x^h B_x)^2}.
\end{equation}
Since for a heavy hole in the thin film regime $g_x^h\ll g_z^h$, the maximal frequency is achieved for a magnetic field perpendicular to the thin film and equals:
\begin{equation}
f_{Ri\bot\max}^h=\frac{\delta x}{h\ell_{so\bot}}|g_z^h| \mu_B B.
\end{equation}
Note that both $f_{Ri\parallel\max}^h$ and $f_{Ri\bot\max}^h$ are $\propto L_z^4$ when $L_z\to0$. In the former case, this results from the $\propto L_z^2$ behavior of $\ell_{so\parallel}^{-1}$ and of the in-plane $g$-factors. In the latter case, this results from the $\propto L_z^4$ dependence of $\ell_{so\bot}^{-1}$, due to the reduced electrical polarizability along $z$.

Eq. (\ref{fRi_perp}) remains valid for light holes. When the gap due to strain is of the order of the gap due to the structural confinement ($\varepsilon_\parallel\simeq2\varepsilon_\parallel^\ast$), the inverse spin-orbit length estimates as
\begin{equation}
\ell_{so\bot}^{-1}\sim\gamma_3(\gamma_3+2\gamma_2)(\alpha_\bot^h-\alpha_\bot^l)\frac{\hbar^2\langle k_y^2\rangle}{(m_0^2/m_\parallel^l)L_z\Delta},
\end{equation}
with $m_\parallel^l$ the light-hole effective mass [Eq. (\ref{m_para_l})]. The light-hole g-factors $g_x^l$ and $g_z^l$ can have comparable magnitudes and the maximal Rabi frequency
\begin{equation}
f_{Ri\bot\max}^l=\frac{\delta x}{h\ell_{so\bot}}\max(|g_z^l|,|g_x^l|)\mu_B B
\end{equation}
is parametrically smaller than Eq. (\ref{fRiparamax_l}). Indeed, $\ell_{so\bot}^{-1}\propto L_z^4$ due to the reduced polarizability along $z$, while $\ell_{so\parallel}^{-1}\propto L_z^2$ increases much faster with small film thickness.

\section{Additional figures}\label{ap:add_figures}

\begin{figure*}
\centering
  \includegraphics[width=0.4\textwidth]{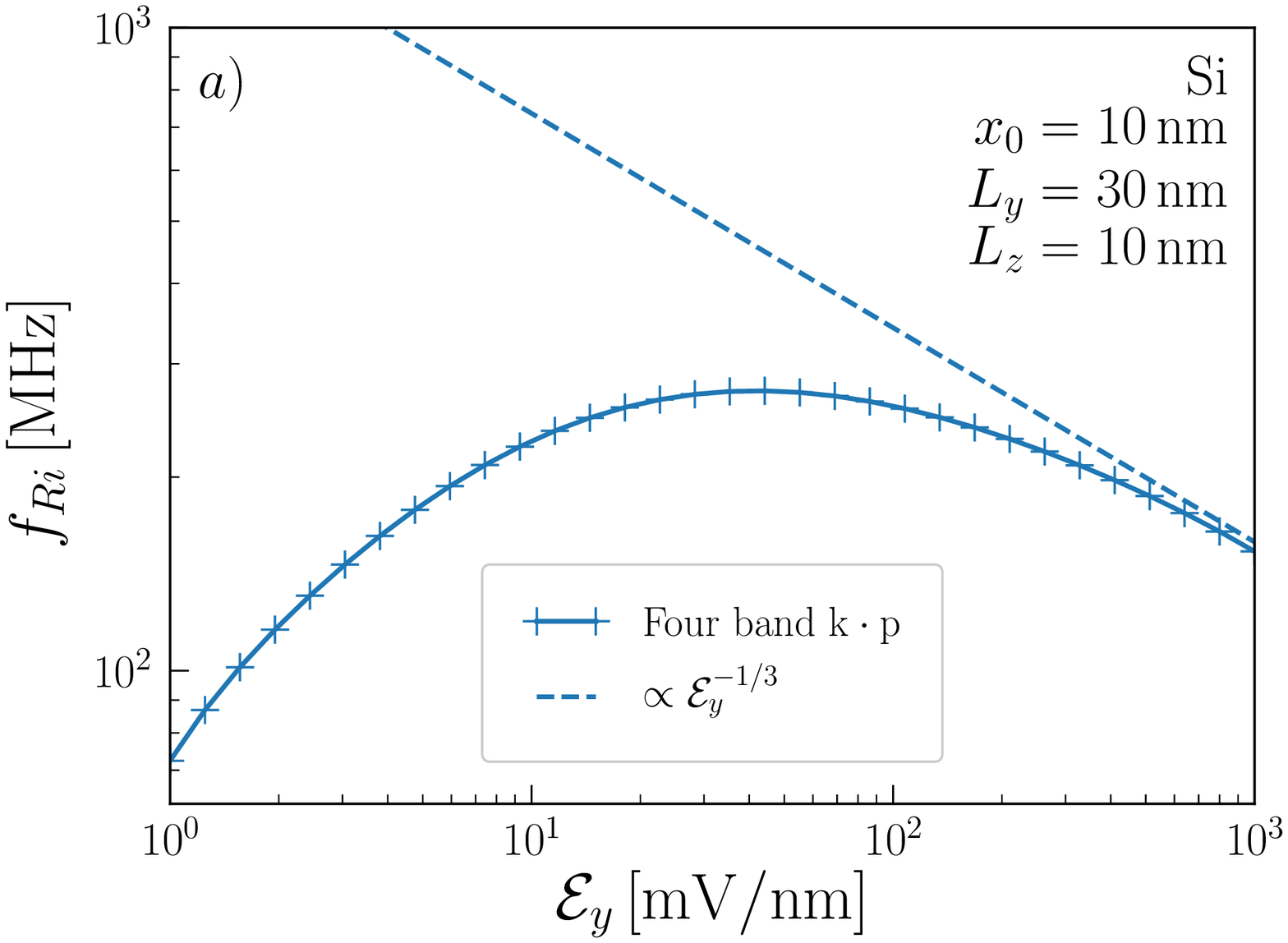}
  \includegraphics[width=0.4\textwidth]{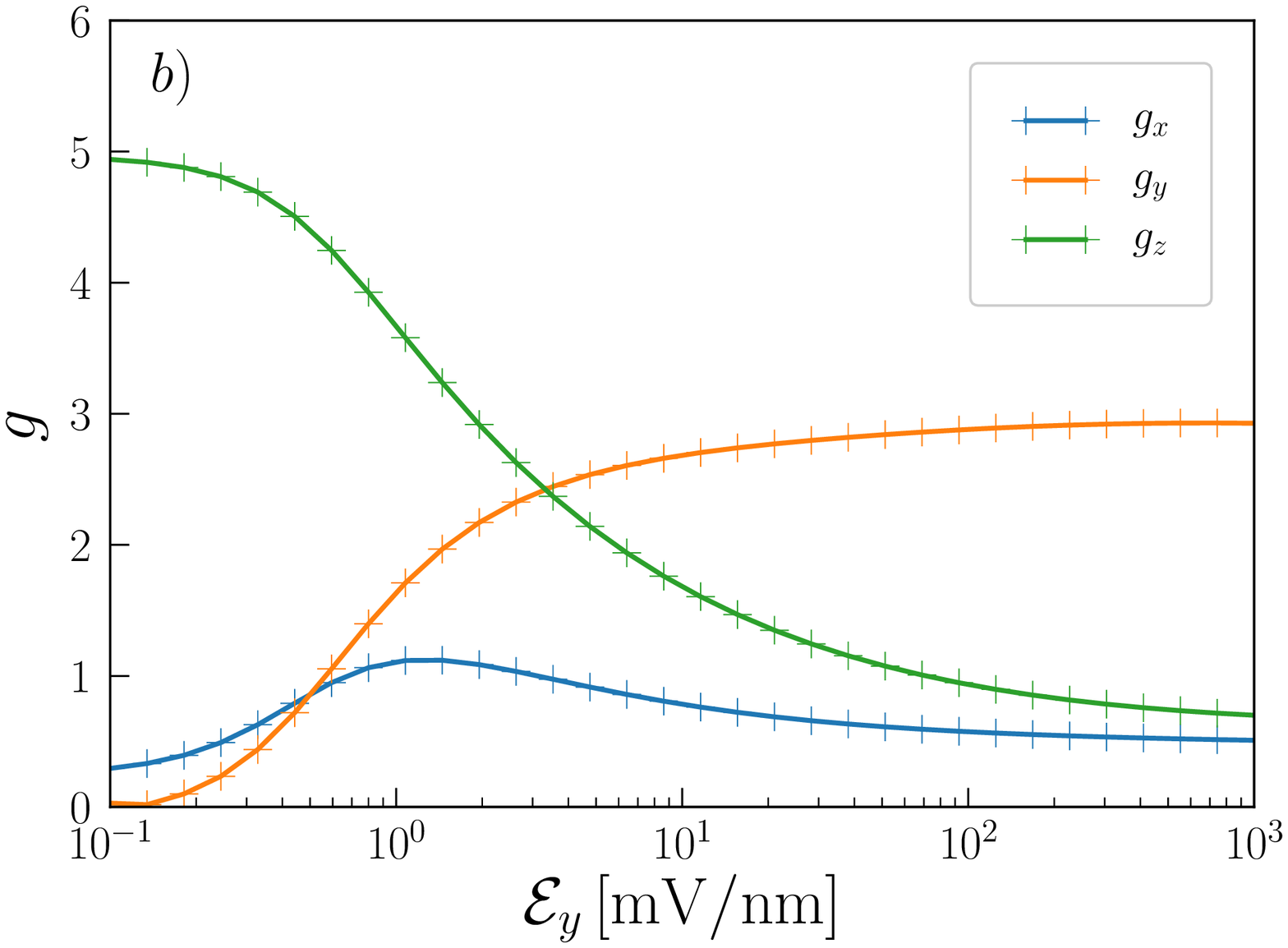}
\caption{a) Double logarithmic plot of the IZ-EDSR Rabi frequency as a function of the static electric field $\ef_y$ along $y$. b) Dependence of the $g$-factors on the logarithm of the electric field $\ef_y$. The data are obtained from the numerical solution of the four-band $k\cdot p$ model\cite{Comment_numerics}. The material is silicon and the relevant lengths are $x_0=10\unit{nm}$, $L_y=30\unit{nm}$, and $L_z=10\unit{nm}$.}
\label{fig:loglog}
\end{figure*}

\begin{figure*}
\centering
  \includegraphics[width=0.4\textwidth]{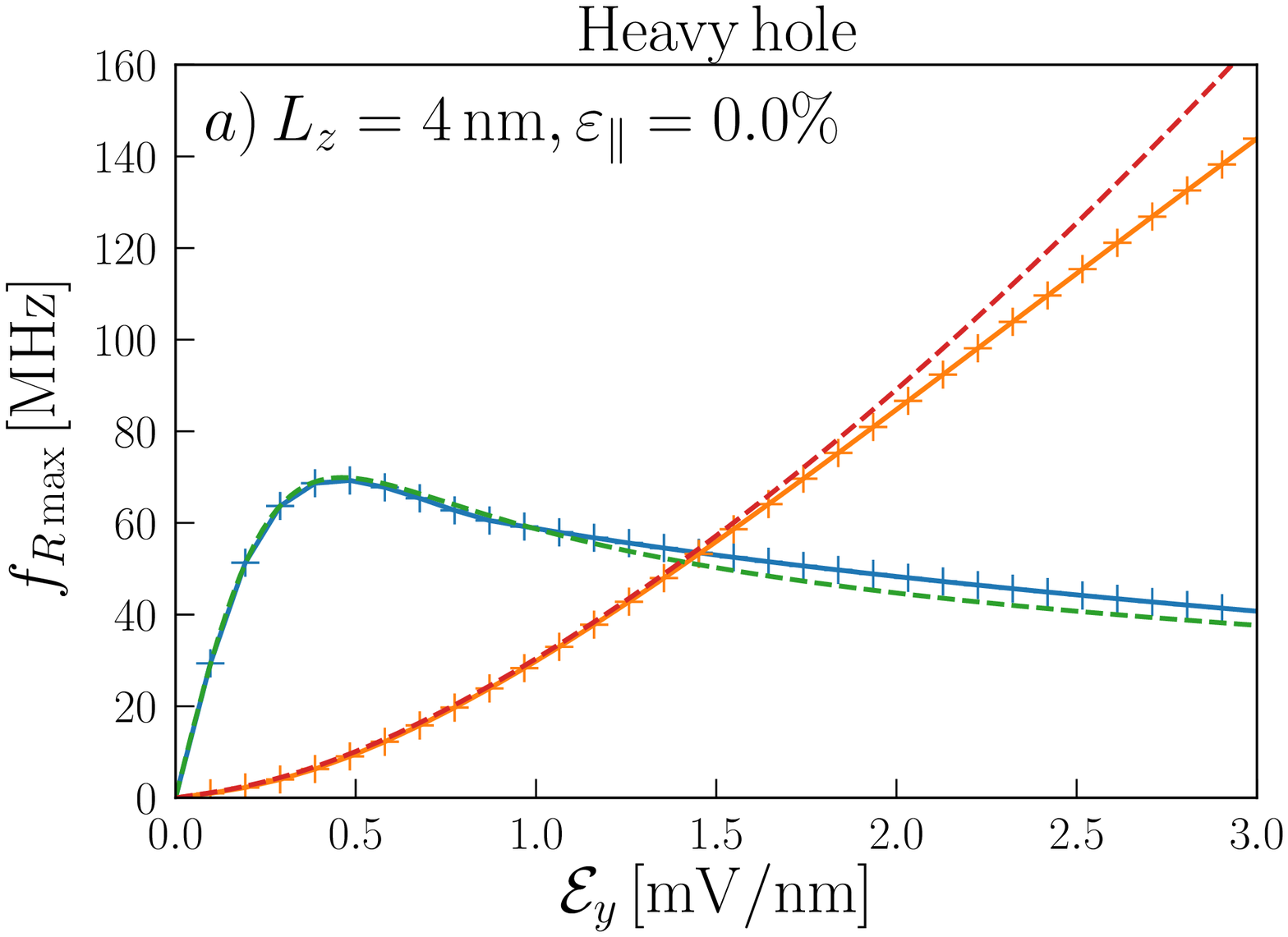}
  \includegraphics[width=0.4\textwidth]{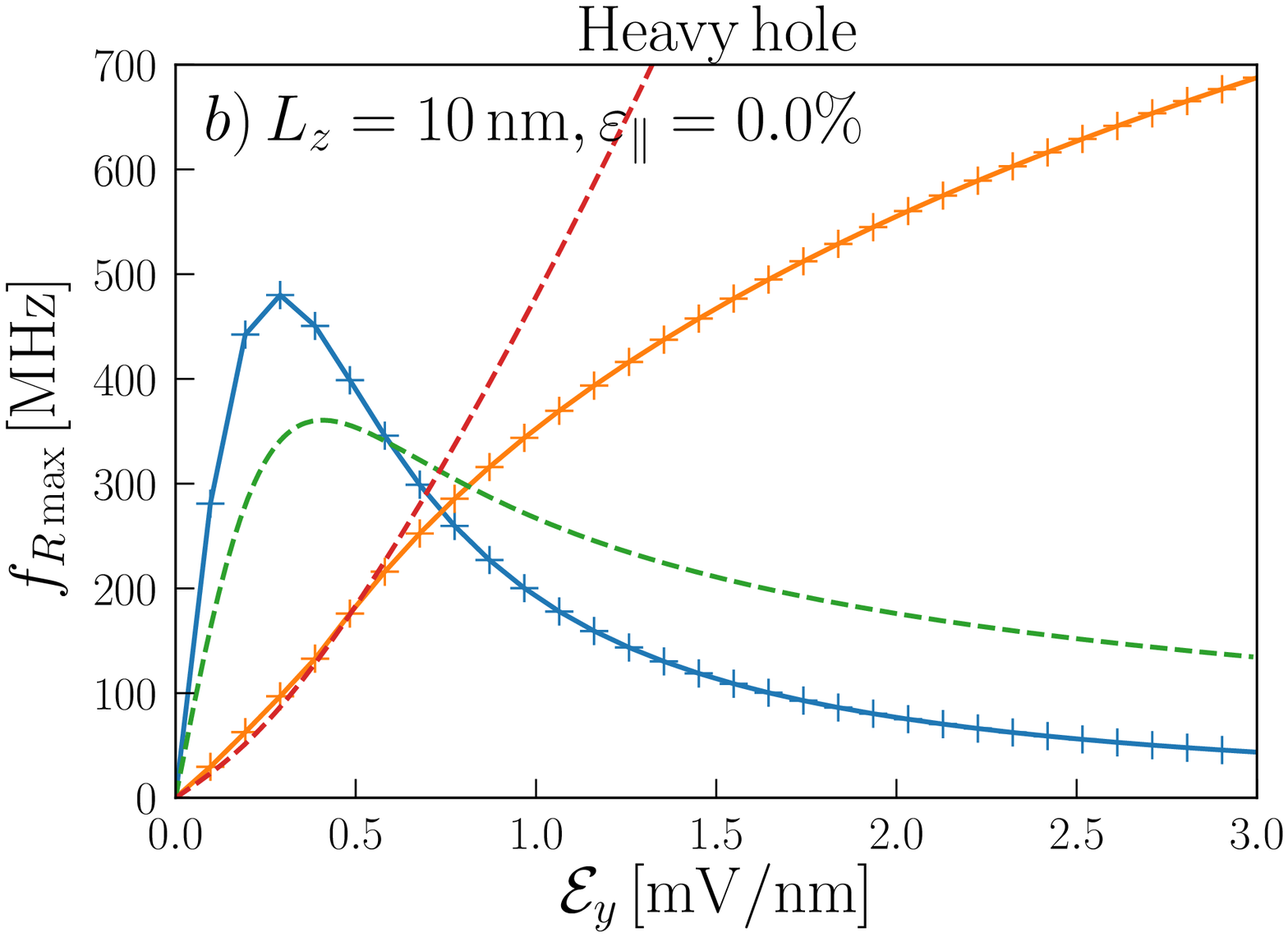}
  \includegraphics[width=0.4\textwidth]{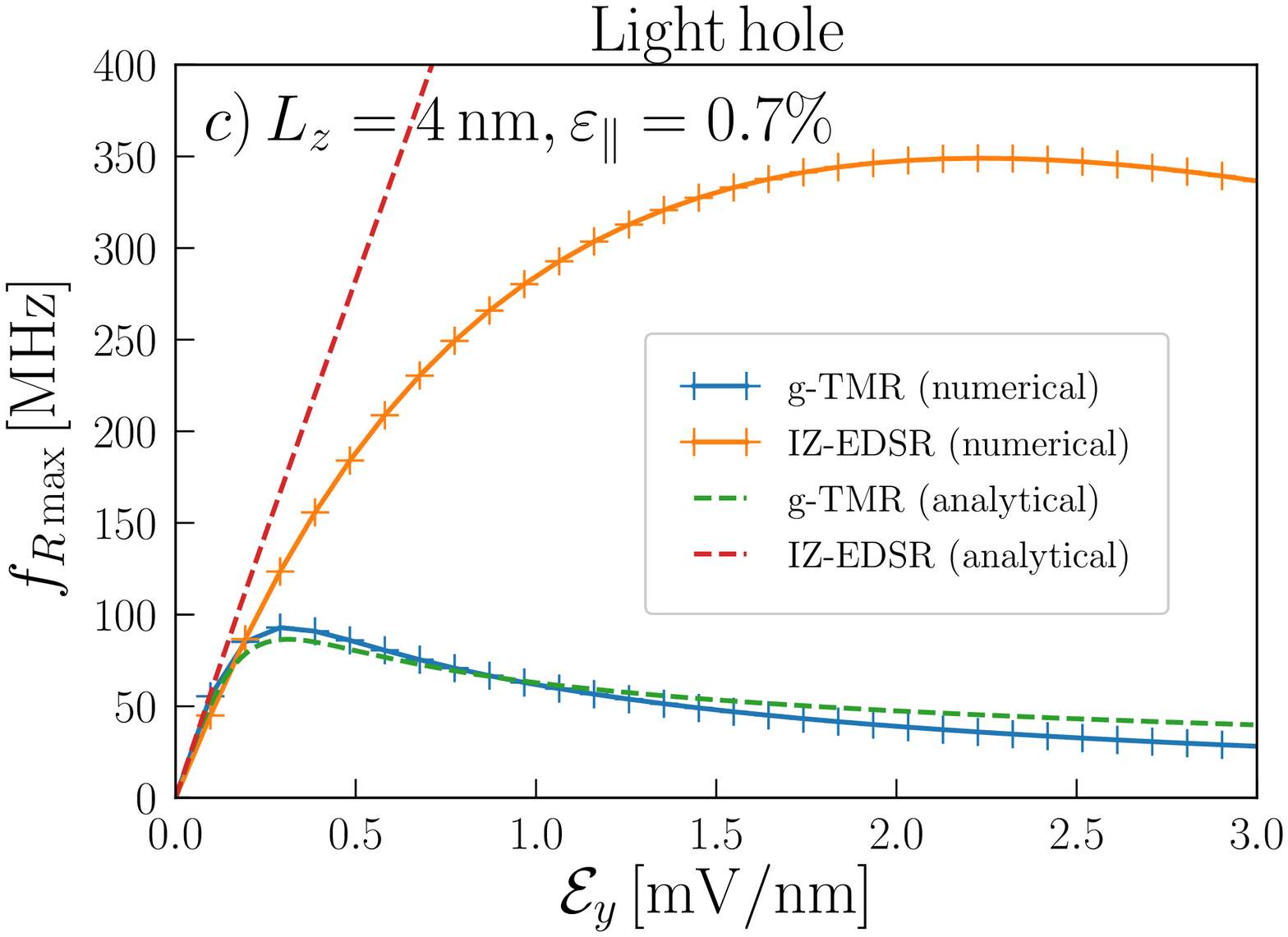}
  \includegraphics[width=0.4\textwidth]{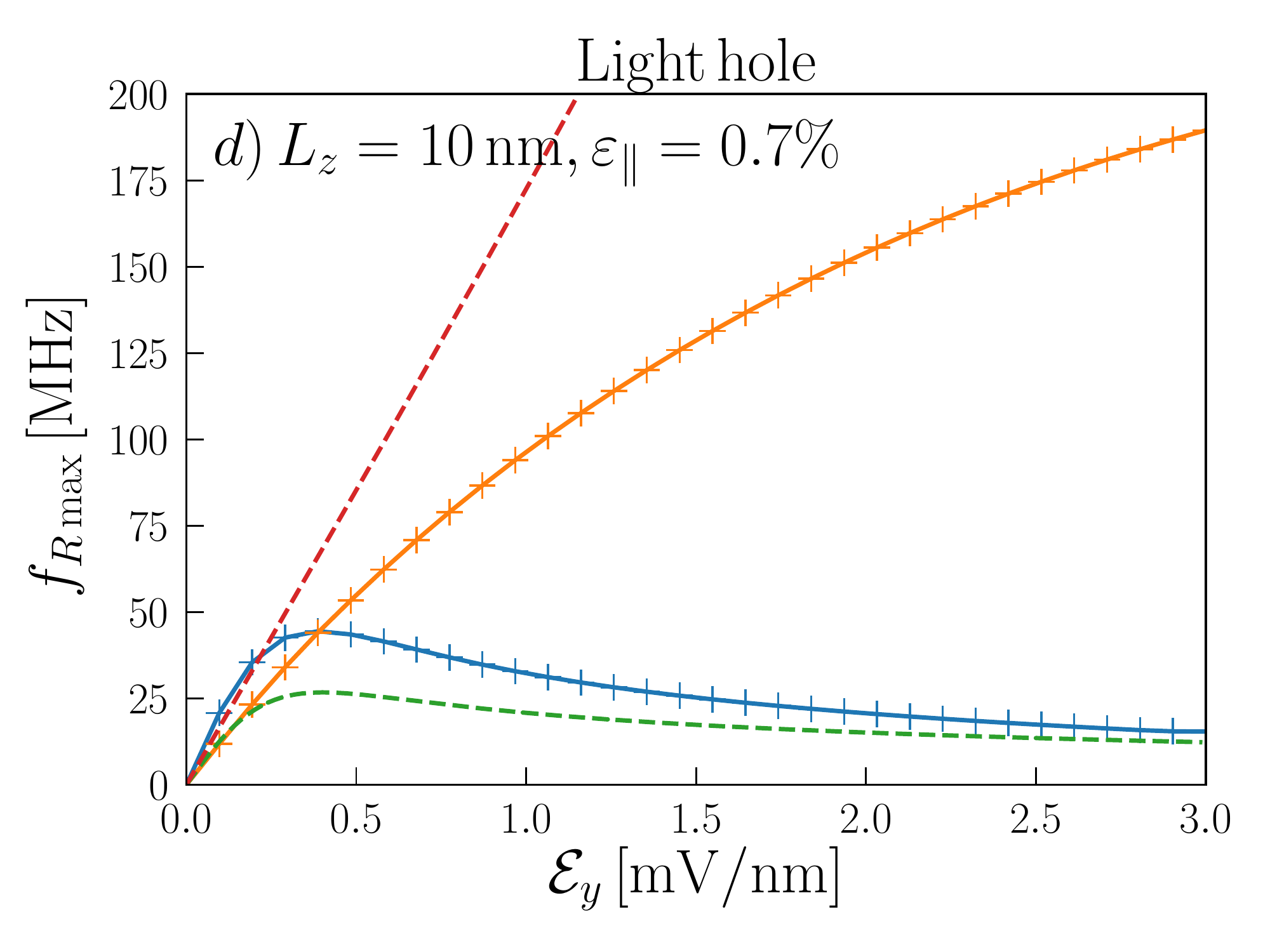}
\caption{Comparison between the maximal $g$-TMR and IZ-EDSR Rabi frequencies of a silicon quantum dot as a function of the static electric field along $y$. The parameters are $L_z=4\unit{nm}$, $\varepsilon_\parallel=0\%$ for figure a), $L_z=10\unit{nm}$ , $\varepsilon_\parallel=0\%$ for figure b), $L_z=4\unit{nm}$, $\varepsilon_\parallel=0.7\%$ for figure c), $L_z=10\unit{nm}$, $\varepsilon_\parallel=0.7\%$ for figure d), $B=1\unit{T}$, $\ef_{x/y}^{ac}=(1/30)\unit{mV/nm}$, $L_y=30\unit{nm}$, and $x_0=15\unit{nm}$.}
\label{fig:Ey_fR_x015nm}
\end{figure*}

\begin{figure*}
\centering
  \includegraphics[width=0.4\textwidth]{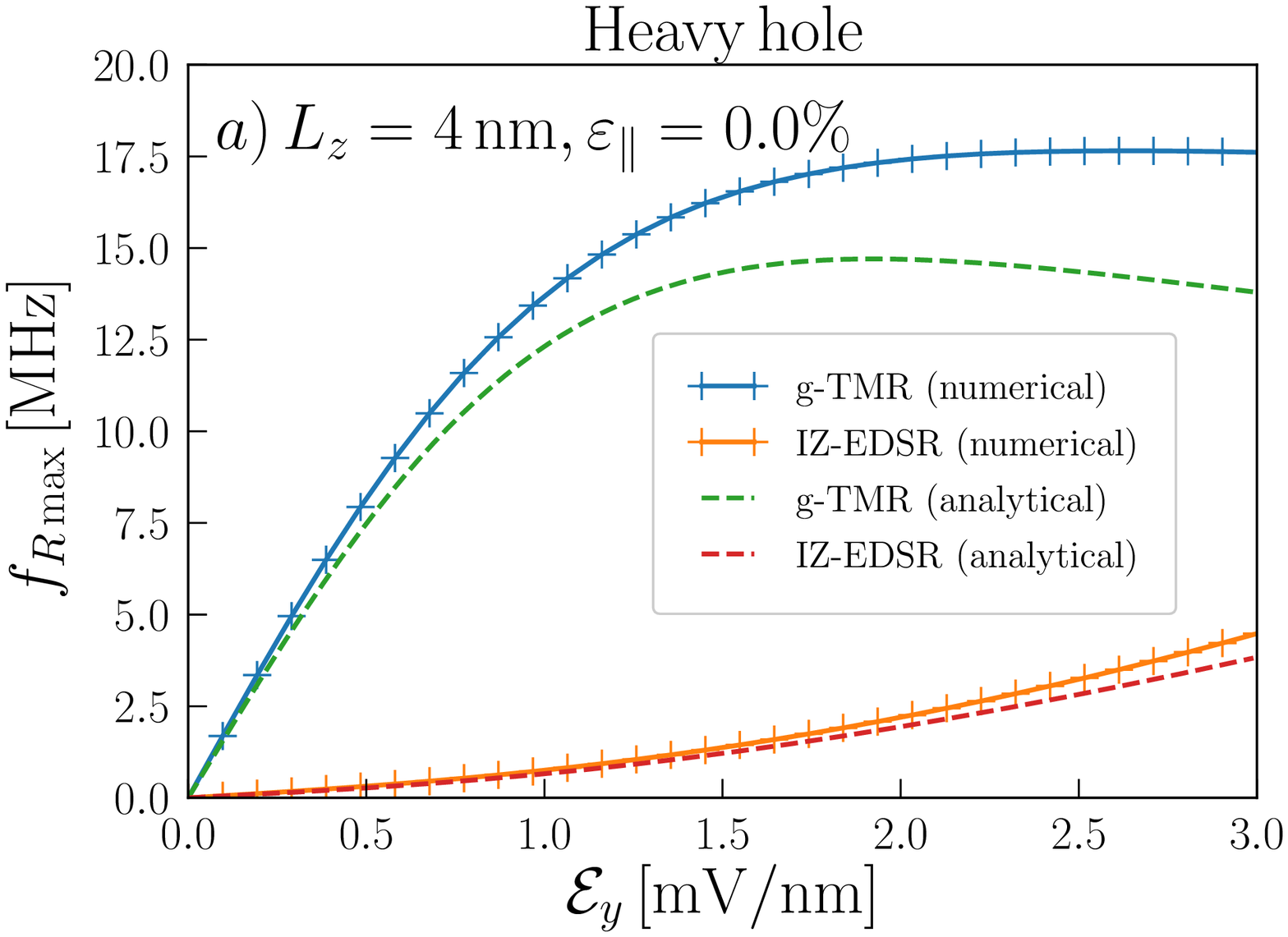}
  \includegraphics[width=0.4\textwidth]{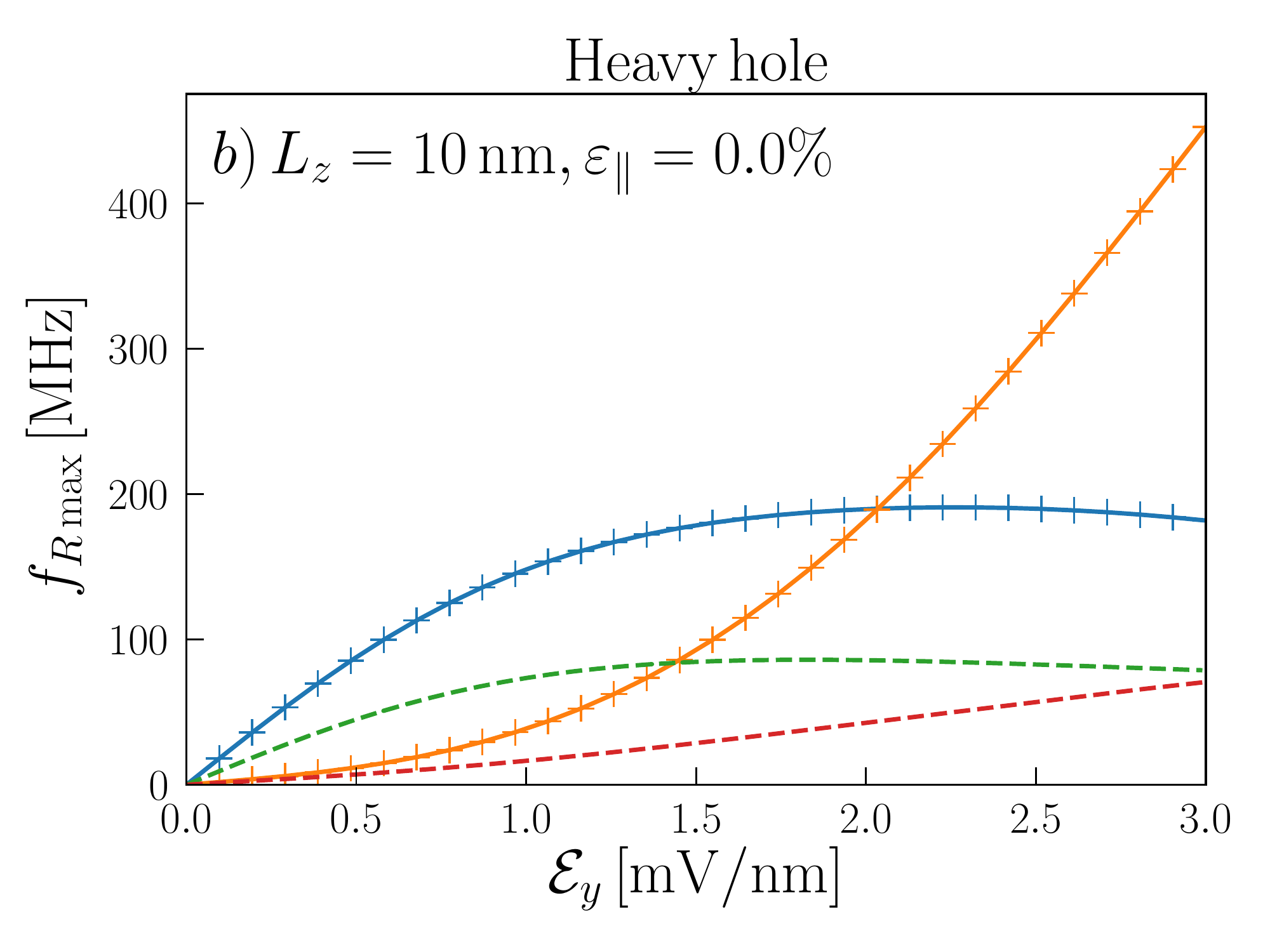}
  \includegraphics[width=0.4\textwidth]{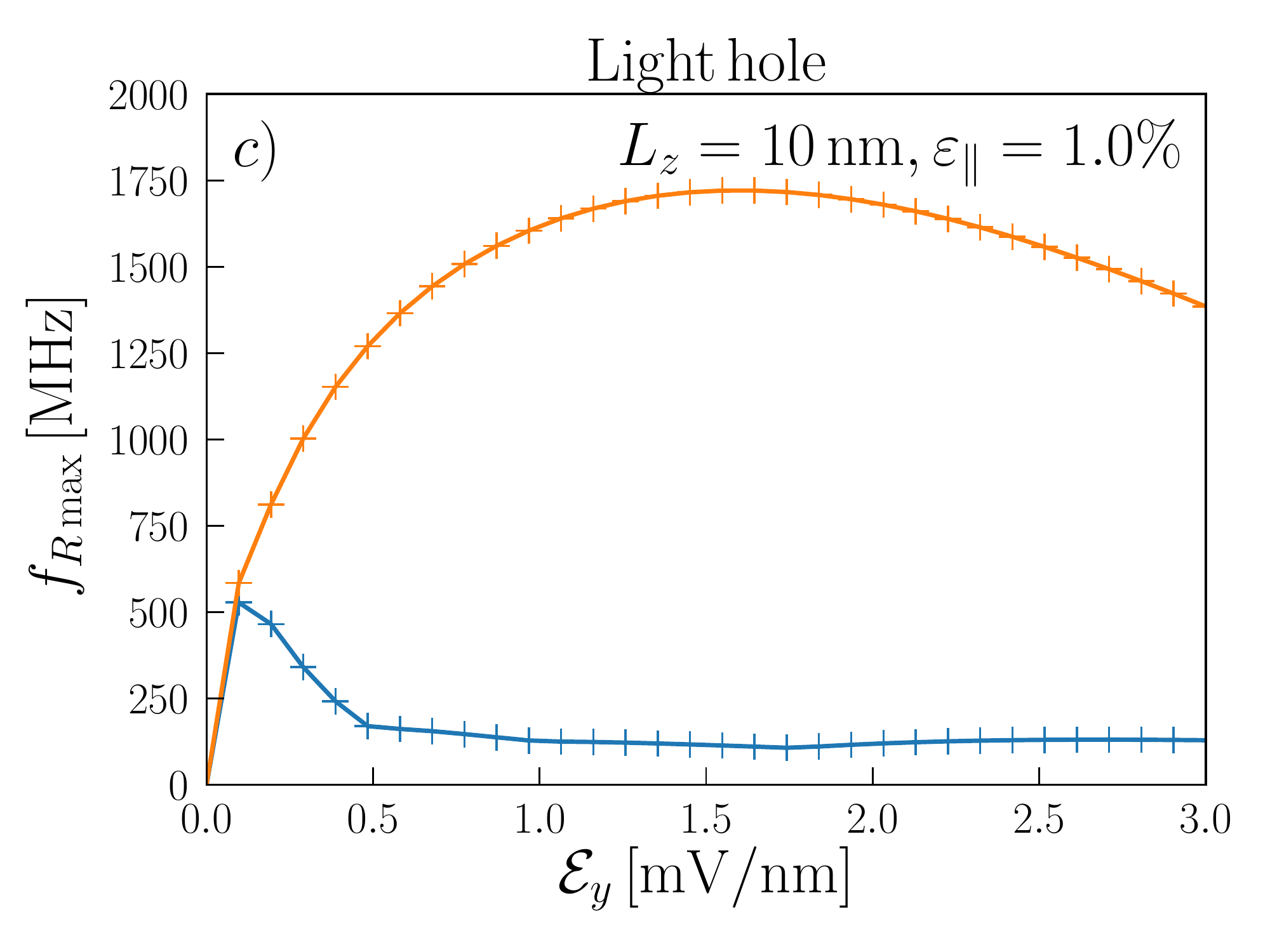}
\caption{Comparison between the maximal $g$-TMR and IZ-EDSR Rabi frequencies of a germanium quantum dot as a function of the static electric field along $y$. The parameters are $L_z=4\unit{nm}$, $\varepsilon_\parallel=0\%$ for figure a), $L_z=10\unit{nm}$, $\varepsilon_\parallel=0\%$ for figure b), $L_z=10\unit{nm}$, $\varepsilon_\parallel=1\%$ for figure c), $B=1\unit{T}$, $\ef_{x/y}^{ac}=(1/30)\unit{mV/nm}$, $L_y=30\unit{nm}$, and $x_0=10\unit{nm}$.}
\label{fig:Ey_fR_ge}
\end{figure*}

\begin{figure*}
\centering
  \includegraphics[width=0.4\textwidth]{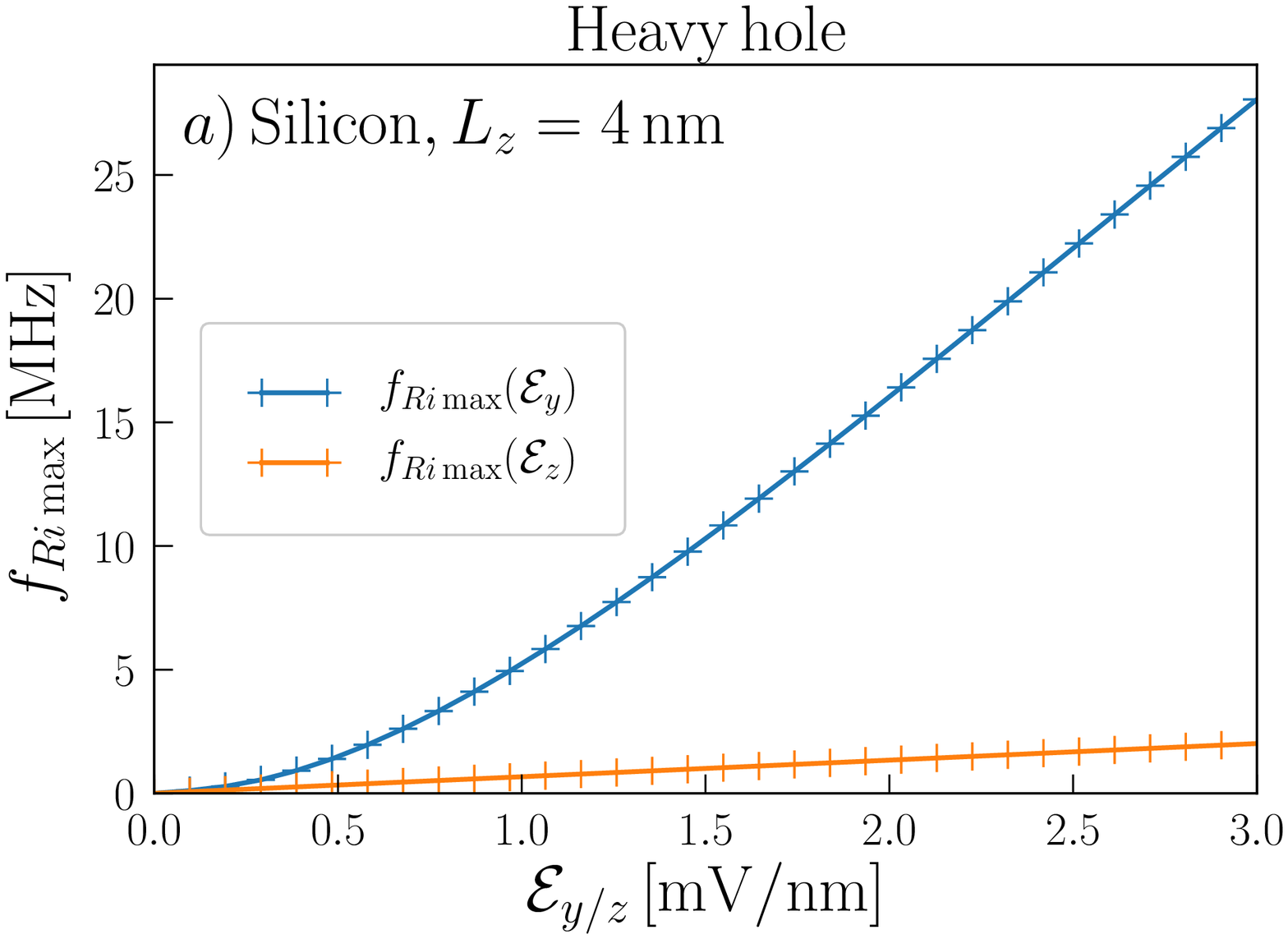}
  \includegraphics[width=0.4\textwidth]{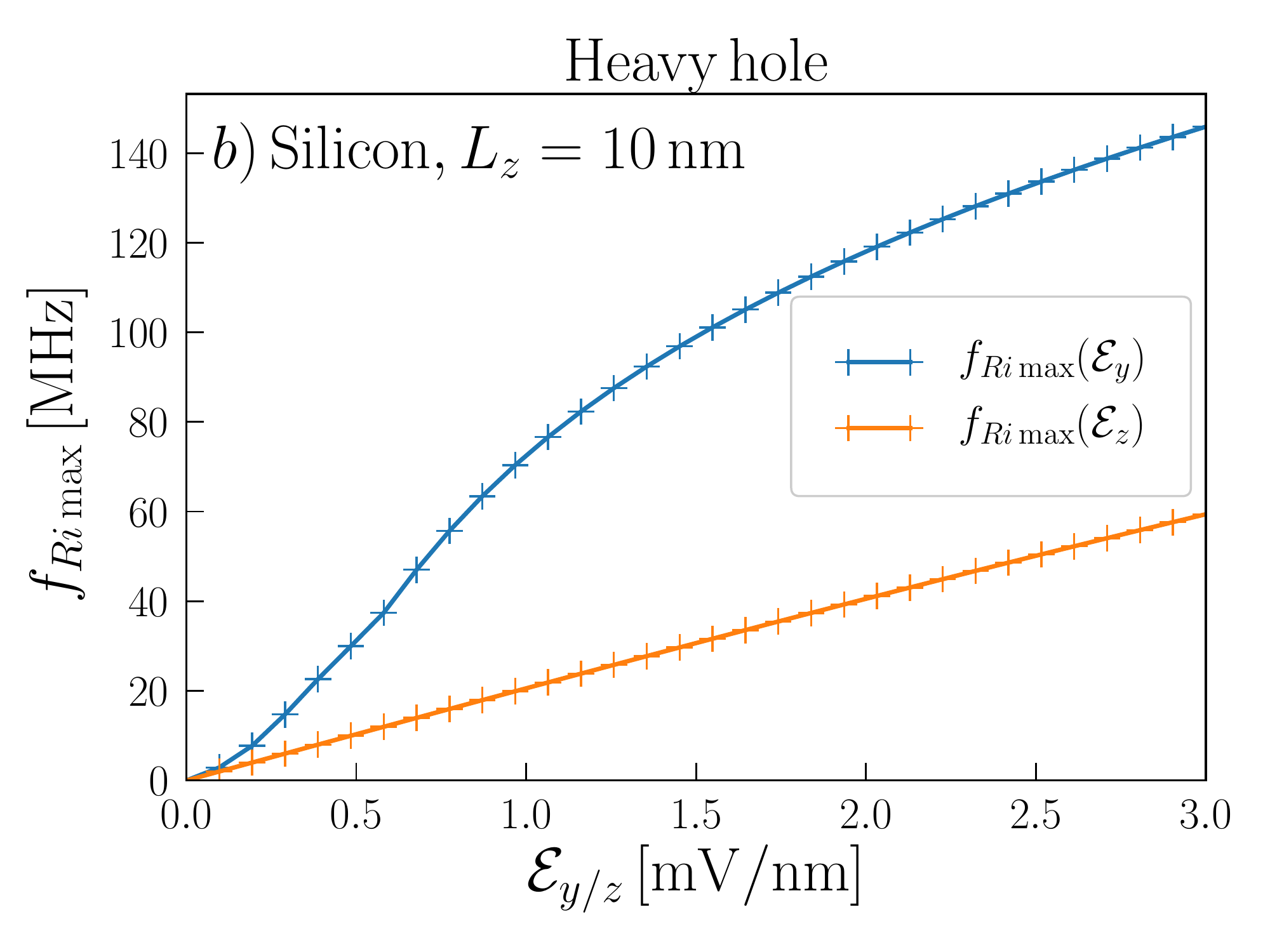}
  \includegraphics[width=0.4\textwidth]{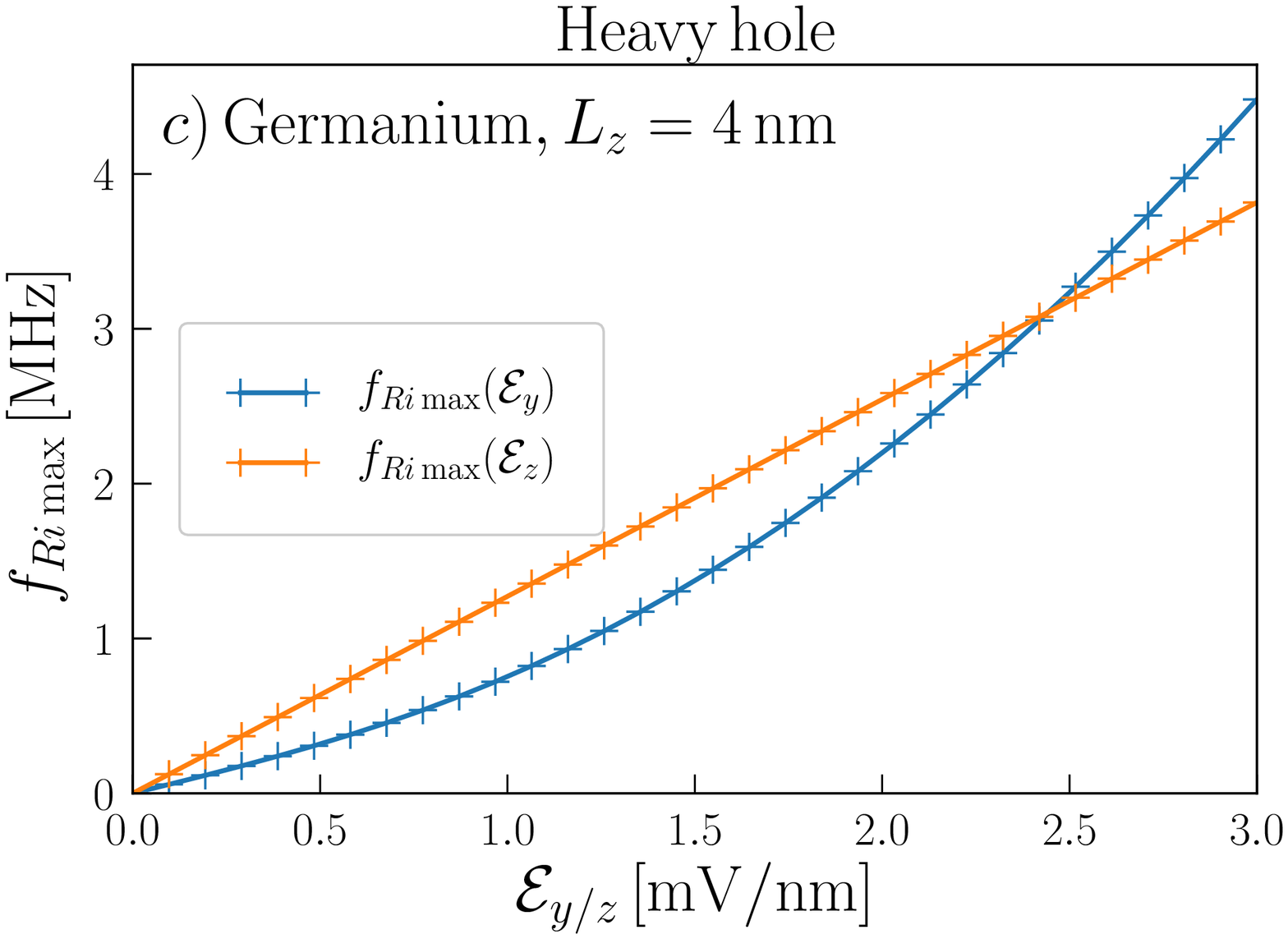}
  \includegraphics[width=0.4\textwidth]{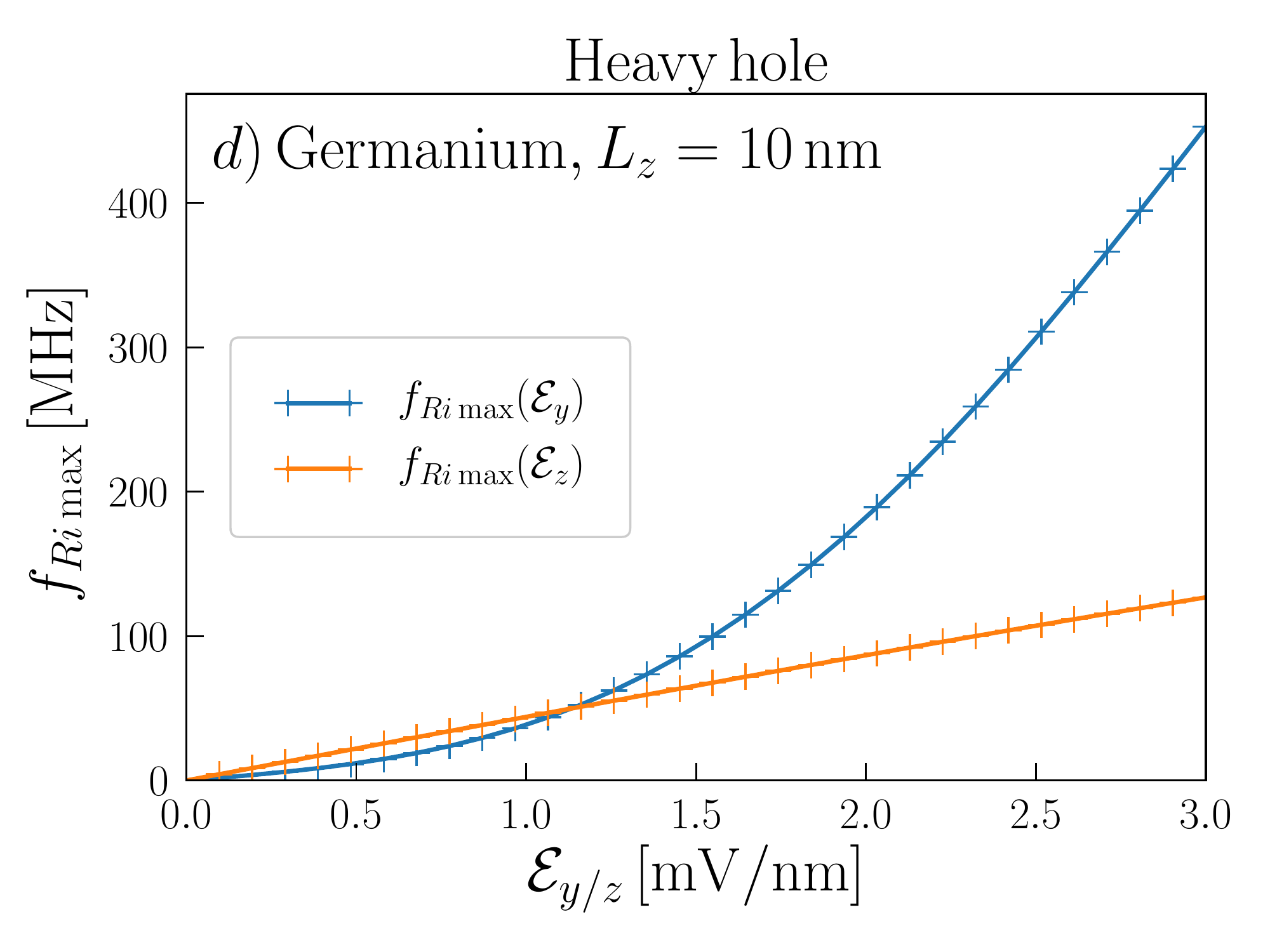}
\caption{Maximal IZ-EDSR Rabi frequency as a function of the static electric field along $y$ and $z$. The parameters are $\varepsilon_\parallel=0\%$, $B=1\unit{T}$, $\ef_{x}^{ac}=(1/30)\unit{mV/nm}$, $x_0=10\unit{nm}$, $L_y=30\unit{nm}$, $L_z=4\unit{nm}$,  for figures a) and c), and $L_z=10\unit{nm}$ for figures b) and d). 
Figures a) and b) are for silicon and figures c) and d) are for germanium.}
\label{fig:Eyz_fRi}
\end{figure*}

We include additional figures. We show in Fig. \ref{fig:loglog} the numerically computed IZ-EDSR Rabi frequency in the strong electric field regime with a highlight on the asymptotic behavior at large field, and the dependence of the $g$-factors on the electric field. In Fig. \ref{fig:Ey_fR_x015nm} we plot the Rabi frequencies of a silicon quantum dot with an increased length of confinement in the direction $x$ of the channel as a support of Sec. \ref{sec:discussion}. In Fig. \ref{fig:Ey_fR_ge} we plot the Rabi frequencies of a germanium quantum dot, also as a support of Sec. \ref{sec:discussion}. Fig. \ref{fig:Eyz_fRi} shows a comparison between the IZ-EDSR Rabi frequencies for static electric fields parallel and perpendicular to the thin film in silicon and germanium. 

\begin{thebibliography}{9}
\bibitem{comment_spin} In the following of the article we use the term spin instead of pseudo-spin, while keeping in mind that for hole qubits the effective spin structure is not only due to pure spin angular momentum, but is also mixed with the hole orbital angular momentum.
\bibitem{Comment_Heff} The basis of the effective Hamiltonian is ($|J=3/2,J_z=3/2\rangle$, $|J=3/2,J_z=-3/2\rangle$) for heavy holes and ($|J=3/2,J_z=1/2\rangle$, $|J=3/2,J_z=-1/2\rangle$) for light holes with $J$ the total angular momentum of the hole and $J_z$ its projection along the quantization axis.
\end{thebibliography}
\end{document}